# Experimental phase transition mapping for hydrogen above 300 K up to 300 GPa

Chang-Sheng Zha[1]*, Hanyu Liu[2], Zhongwu Wang[3], William A. Bassett[4]

[1]Earth and Planets Laboratory, Carnegie Institution of Washington, 5241 Broad Branch Rd. N.W., Washington DC, 20015, USA
[2]International Center for Computational Method & Software, and State Key Lab of Superhard Materials, College of Physics, Jilin University, P. R. China
[3]Cornell High Energy Synchrotron Source (CHESS), Cornell University, Ithaca, New York 14853, USA
[4]Department of Geological Sciences, Cornell University, Ithaca, New York 14853, USA

**Abstract** A vast amount of Raman spectroscopic data, obtained using diamond anvil cell (DAC) technique under *in situ* high pressure-temperature (PT) conditions, has been used for mapping the phase transitions of hydrogen in the temperature range of 300 – 900 K and pressure up to 300 GPa -- a PT region in which most phase information was unknown previously. The isothermal and isobaric dependencies in frequency and peak width for intramolecular vibration Raman mode of hydrogen (vibron $\nu_1$) are established based on thousands of data files which are measured during hundreds of independent high PT experimental runs. Discontinuities of pressure/temperature dependencies in frequency/peak-width, together with appearance or disappearance of observed Raman modes were obtained. Those transition data show self-consistency and have been used for outlining 13 phase transition boundaries and 15 possible phases. This large, diversified behavior of spectra demonstrates that surprisingly rich and complicated phase transitions may exist in this PT region.

## Background

Precise prediction of hydrogen phase change including metallic transition has been a long standing effort because of the enormous zero-point energy (ZPE) for this lightest element.[1] The large quantum fluctuation strongly affects the theoretical calculations such that there has been continuous improvement in the predicted transition pressure for metallic hydrogen.[2-4] Experimental methods are still the only reliable ways to study the phase diagram of the hydrogen system. However, enormous technical challenges exist for studying this simple element especially under extreme pressure-temperature (PT) conditions. In the last several decades, only four solid hydrogen phases have been confirmed by experiments, which were mostly conducted below or at room temperature.[1,5,6] A large portion of the phase diagram above room temperature up to the melting curve still remains unknown. There are a few experiments with limited PT data points above 300 K that have been tried but, the reported results were far from convincing and not self-consistent.[7] In this study, we report the results of a systematic phase mapping experimental effort, based on thousands of *in situ* high temperature Raman scattering data from room temperature to the melting curve covering a pressure range up to 300 GPa, in which we reveal surprisingly rich phase transition behaviors for this almost unknown PT region for hydrogen.

X-ray and neutron scattering serve as the most effective tools for determining material structures and phase transitions. However, the low scattering cross section for hydrogen and small volume when it is under extremely high pressure, make hydrogen almost invisible to these methods. Alternatively, vibration spectroscopies like Raman scattering and IR absorption are more effective and mostly preferred in studying the high PT phase transitions of hydrogen.[8-16] The activity of Raman mode depends on the lattice symmetry, therefore, phase transition induced by the changing of two physical variants -- pressure or temperature -- should cause Raman modes to appear, disappear, or undergo discrete frequency/peak-width shifting. There are two phase mapping pathways that can be used for finding these physical discontinuities: isothermal compression can obtain the pure pressure dependence of the Raman modes, or isobaric temperature variation process for obtaining the pure temperature dependence of Raman modes. Discontinuities in either of these two dependencies are considered to be the possible phase transition points at the phase boundaries. In principle, the important prerequisite is that either pressure or temperature must be varied while the other one is kept constant. Most previous PT experimental studies investigating the phase transitions for hydrogen were conducted at or below room temperature, using either of the above methods with relative ease for keeping one of the two variants constant. This makes the conventional "reversal" procedure, in which pressure or temperature is varied back and forth while keeping the other one constant, is practical and actually the common method for determining phase transitions. However, heating hydrogen above room temperature up to ~1000 K under very high pressure causes big technical challenges for the experiment, because the hot dense hydrogen has a strong tendency to break diamond and damage the gasket materials. This, on the one hand, requires the smoothest and shortest possible heating durations, which can greatly reduce the possibility of chemical reaction between hydrogen and diamond or gasket. On the other hand, changing the force-loading to the DAC during heating often introduces a high rate of diamond/gasket failure. Therefore, isothermal pressure adjustment at constant high temperatures is not preferred. Heating the sample while maintaining a constant force-loading can greatly improve the chance that the sample will survive to higher temperatures even at very high pressures. For this reason, we chose to use the procedure of heating at a constant force-loading as the preferred method for each run.

In a previous study, we have successfully used the externally heated diamond anvil cell in conjunction with the constant force-loading heating method, combined with micro-Raman scattering system for studying the melting curve of hydrogen.[17] High quality Raman spectra for intramolecular vibration (vibron) and lattice modes can be obtained up to 300 GPa at 295 – 900 K, where the disappearance of lattice modes can be used as the melting criterion at certain pressure-temperature points, that define the melting curve. However, the same technique for discriminating solid-solid phase transitions is more complicated because the transition criteria are no longer unique as with melting. Except for the mode appearance or disappearance, many phase transitions may be manifested only with a discontinuous shift of mode frequency. Unfortunately, heating process can always cause spurious random pressure change even though the DAC's force-loading mechanical parts are not touched. These may result from the complex effects of thermal expansion of DAC parts and deformation of the metal gasket. These changes can also produce Raman frequency shifts. Sometimes, partial sample leakage could be mixed with the changes mentioned above making relatively larger pressure drop and corresponding frequency change. So, during a constant force-loading heating run, frequency shifts from both temperature and pressure change will be mixed and can be impossible to distinguish in order to recognize each contribution from those changes. To avoid this confusion, making loading-force change during heating to compensate for the pressure change was considered but proved to be too difficult and too unreliable. Compensation adjustments are very hard to control, especially because it causes a high rate of diamond/gasket failure similar to those caused by isothermal compression at high temperature. Figure 1 shows example experimental runs which demonstrate the complications described above. The confusion in physical origin of Raman mode changes makes judgment of phase transition from the raw frequency-temperature data, obtained in a single constant force-loading heating run, impossible.

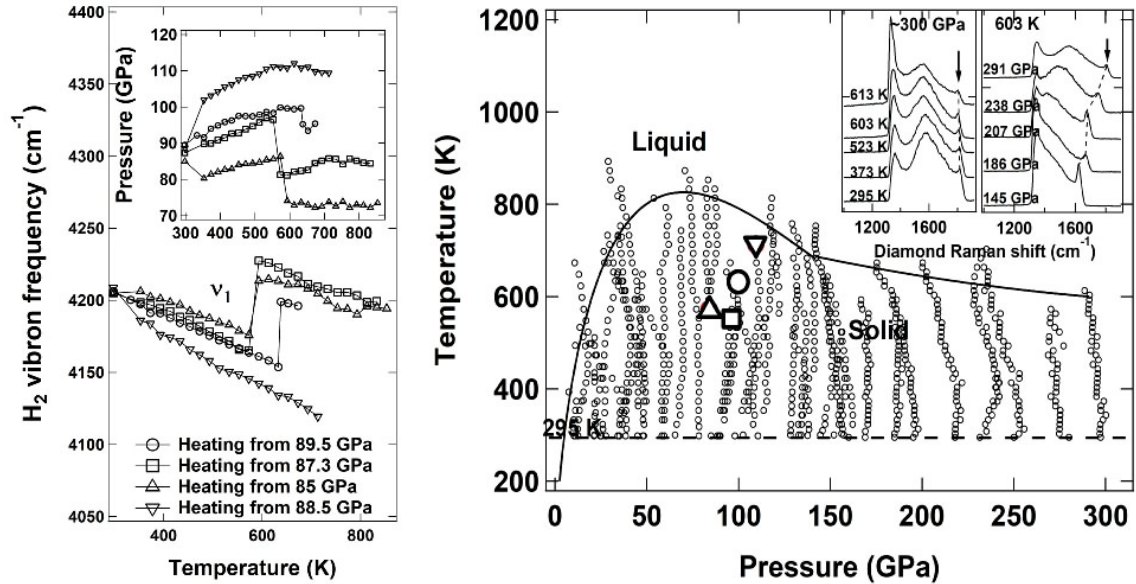

Figure 1. Left: Frequency-temperature dependence of hydrogen vibron $\nu_1$ at four constant force-loading heating runs with similar starting pressures. The random frequency discontinuities accompanying large random pressure changes (inset) during the heating make difficulties for determining its physical origin. Right: Plot of partial experimental data points (circles) and melting line (solid line), *in-situ* measured pressure and temperature information of the points have been used for discriminating phase boundary in this study; the inset shows diamond Raman spectra with sharp peak in the high frequency side were used for in-situ pressure determination. Three of four large different symbols (triangle, square, and circle) of right panel indicate the three PT locations of the frequency discontinuity, each with the same symbol as the corresponding heating run shown in the left panel respectively. The downward-pointing triangle symbol shows the end pressure-temperature location of fourth run, in which no pressure and frequency discontinuities were seen. Random pressure trajectories during each constant force-loading heating run cause different locations of frequency discontinuity in the PT diagram though their starting pressures were similar, and those locations can not be used for determining phase transition directly (see text for details).

We need to distinguish the mixed situation for finding the pure effects resulting from pressure and temperature. In order to accomplish that, a large amount of PT data points with dense temperature and pressure steps are needed to organize data into an isothermal and isobaric vibron frequency network. Such an approach should permit identification of the discontinuity points found on each dimension of the network in order to locate the phase boundaries.

## Experiment and results

In an almost decade long period, hundreds of runs of high PT experiments for hydrogen have been conducted.[17] About ~3500 spectra from more than a thousand PT data points were obtained initially for the purpose of detecting melting (Figure 1). In each heating run, a fresh high-pressure sample was loaded and the sample was compressed to a desired pressure at room temperature, then paused until the pressure stabilized with final pressure checked by the vibron frequency at room temperature published previously.[15] A pre-set stepwise heating under constant force-loading was performed. High quality *in situ* pressure measurements during heating are critically important, especially at very high pressure. A sharp peak feature in the high frequency side of broadened diamond Raman spectrum (right panel inset in Figure 1) is always obtained even at very high pressure and temperature. It is sharp and well defined because the relatively soft nature of hydrogen causes the pressure condition to be quasi-hydrostatic.[9,18] Spectra have demonstrated that the intensity of this peak is proportional to hydrogen sample volume;[15] a weaker peak corresponds to a smaller sample volume and disappearance of a peak most likely indicates some source other than hydrogen. The position of the sharp peak is sensitive to pressure in the heating run and was used for the *in situ* pressure correction. The uncertainty for locating the peak position is $\pm\, 2$ cm$^{-1}$. The left and right insets at the right panel of Figure 1 show example spectra of diamond Raman in a constant force-loaded heating run of the highest starting pressure, and a very high-temperature isothermal plot respectively.

The *in situ* pressure and temperature values for each data point are obtained as described in the "Method" section. Although a single heating run does not offer the correct phase transition information as discussed above, a systematic reanalysis of the whole data set can provide a correct solution. The first step is to create the isothermal pressure dependency of the vibron frequencies, using the same temperature points obtained from each of the constant force-loaded heating runs. Those runs cover a wide PT range up to ~300 GPa and ~900 K. Figure 2 shows examples of isothermal pressure dependencies of vibron $\nu_1$ frequency and peak width at four different temperatures. The findings of vibron frequency first decreasing with pressure above ~30 GPa, and having a significant slope change around ~216 - 220 GPa at room temperature have been well established experimentally.[9,10,15] At higher temperatures, we found that the starting pressures for the significant slope change vary with temperature and that the isothermal curves show much more complicated details, indicating it is crossing over many different phase boundaries as shown in Figure S1 in the supplementary materials (SM). Since the data points come from different independent heating runs, they show small random, unavoidable scattering (standard deviation $\sigma = \sim 4$ to $\sim 10$ cm$^{-1}$). However, statistically smoothed sectional tendency between obvious larger discontinuous points ($\Delta = \sim 20$ to 45 cm$^{-1}$), indicates possible phase changes. The isotherm derived in this way has the advantage of reducing possible systematic pressure-frequency errors that could be caused by a single isothermal compression experiment. This random averaging by a statistical probability might help to reveal a few very small discontinuities such as found at the point, for example, at ~150 GPa and 300 K, which has not been reported in most previous room temperature compression experiments except for one.[19] It is not surprising that the majority of pressure-frequency discontinuities, if not all, appear simultaneously with that of peak width at the same PT conditions as shown in the plots (Figure 2 and S1 of SM), this self-consistency provides evidence for the reliability of the measurements. Solid lines in Figure 2 are the best fit to the smoothed sections separated by the frequency discontinuities. These lines offer one way of sorting out isobaric temperature dependence of vibron frequency at different pressures.

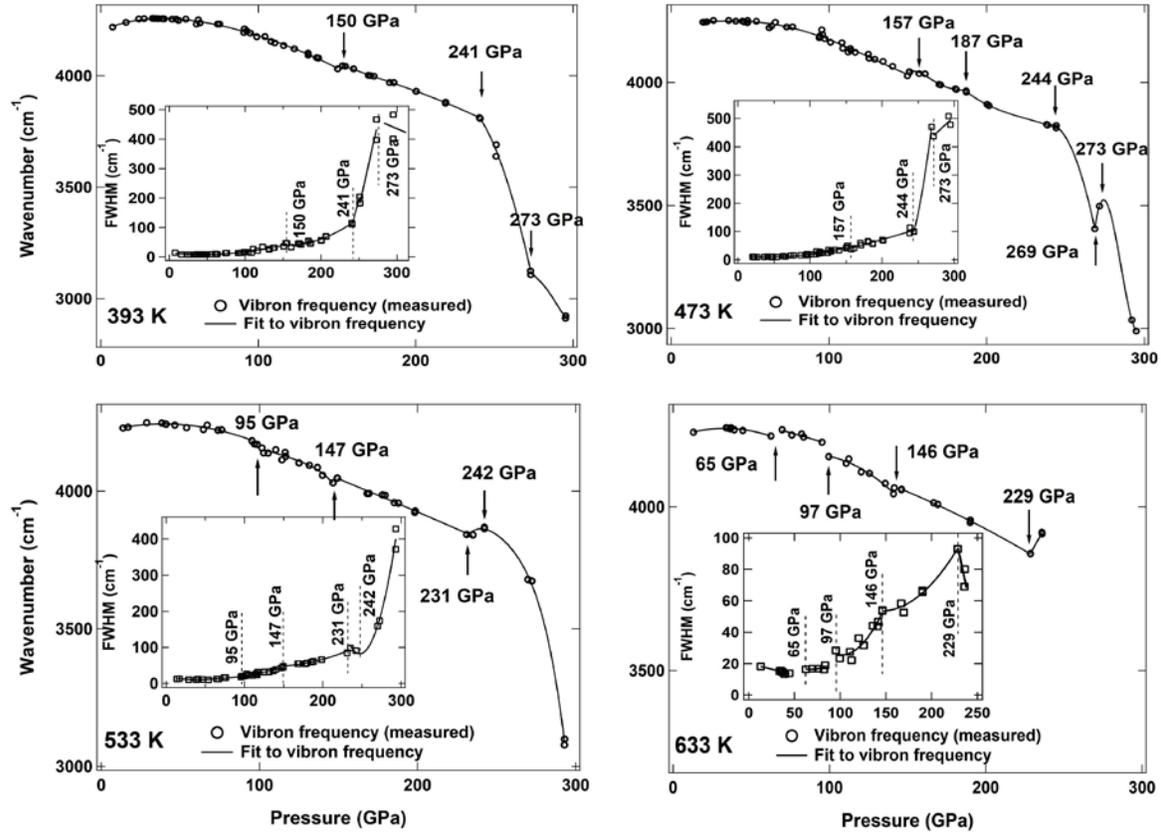

Figure 2. Plot for isothermal pressure dependences of vibron frequency and peak width (insets) at selected temperatures. Symbols are data points; each of them comes from an individual heating experimental run. Small arrows mark kinks in vibron frequency; vertical dashed lines mark kinks in FWHM. Solid lines within each kink-free interval are best fitting curves for that pressure range, which offered a possibility for calculating frequency at desired pressures and constructing isobaric plots. Twenty-seven isothermal plots covering the range 295 to 813 K are listed in supplemental information.

The second step is to establish isobaric frequency-temperature (FT) relations based on the fitted isothermal frequency-pressure (FP) dependences. Figure 3 presents combined plot for the temperature trend of the $v_1$ frequency obtained during selected constant force-loading heating runs and the isobaric corrections using the isothermal fitting results. The measured peak width (FWHM) variations for vibron $v_1$ are also shown as insets. Some isobaric plots clearly reveal significant FT discontinuities which were not observed during constant force-loading heating runs. It is because the accompanying pressure change modified the whole heating frequency trajectory. Similar to the isothermal plot, the consistency between discontinuities of frequency and peak width is observed for many transitions. Temperature dependencies of vibron frequency also reveal some variations that are rarely seen in pressure dependence. For example, we observed that negative FP slope almost dominates the changing tendency in most pressure ranges except one phase area, a unique behavior for phase I'$_{hb}$ (Fig. 4), it shows a positive FP dependence for vibron $\nu_1$ (see Fig.S1 of SM #11 - #18). In contrast, FT dependence shows complicated fluctuation in slope at different PT range (see Figure S2 of SM).

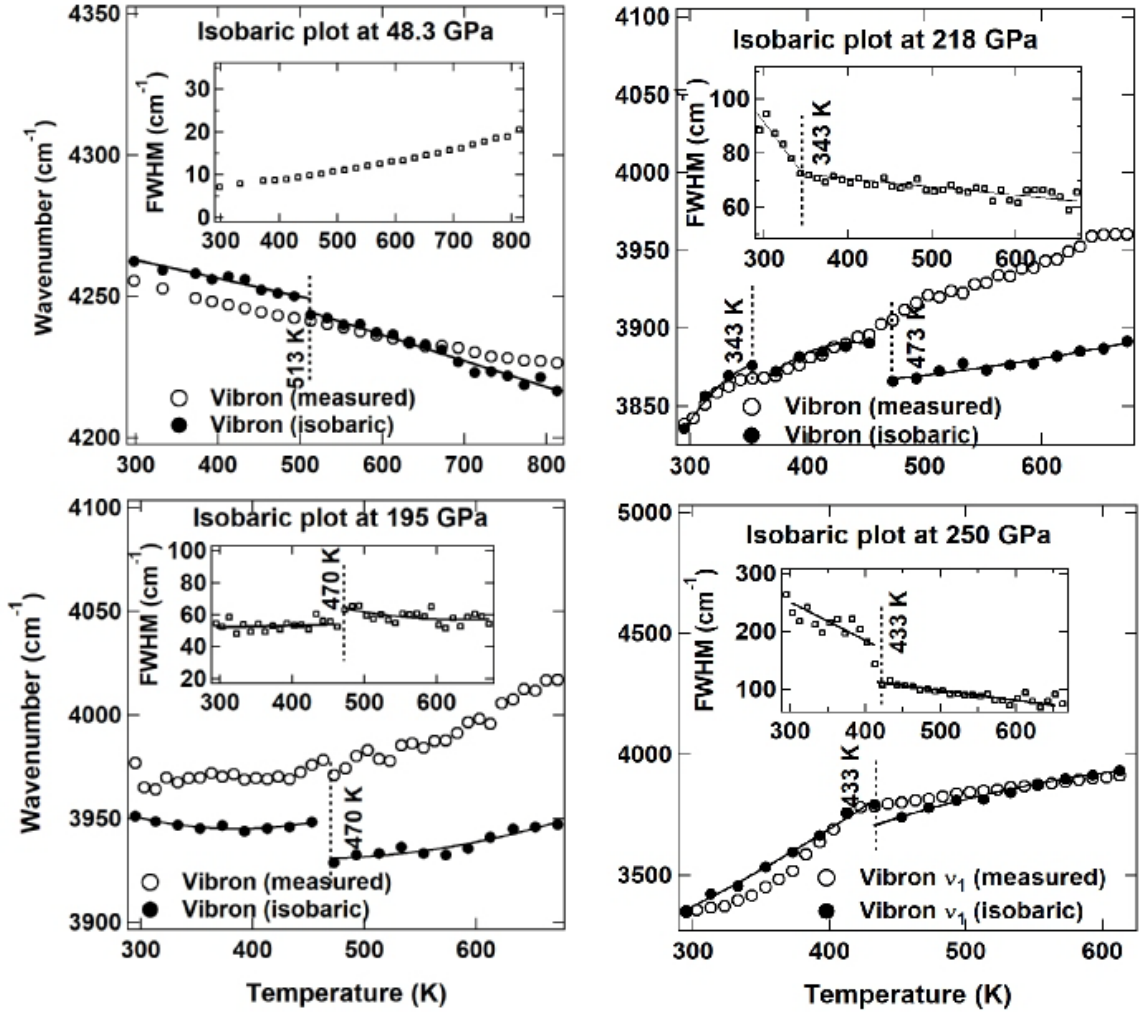

Figure 3. Isobaric frequency-temperature plots for vibron $\nu_1$ at four different pressure runs. The frequencies measured during the constant force-loading heating run were under random pressure conditions, whereas the isobaric vibron frequencies are obtained from the fitted isothermal curves at different temperatures as shown in figure 2 and S1. Symbols are the data points; solid lines are for eye guiding only. Plots for peak width changing with temperature are also shown with inset.

A total of 54 isothermal and isobaric plots were obtained and listed in the supplementary materials. All the P – T points showing discontinuous frequency, peak width, and slope changes found in either of these two plotting categories were chosen to define the possible phase boundaries. Figure 4 represents the plot of all those transition PT points.

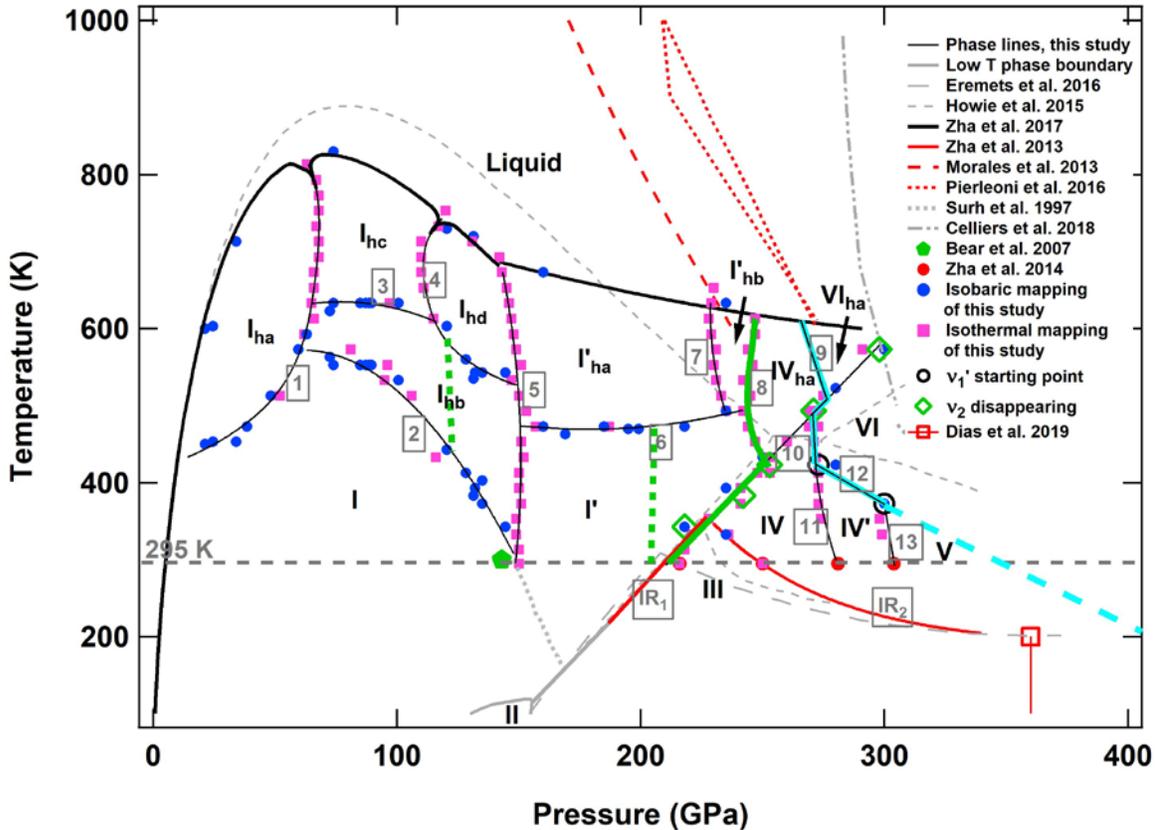

Figure 4. Phase diagram of hydrogen. Pink squares and blue solid dots are the phase transition points from isothermal and isobaric mapping respectively. Black thin solid lines are phase boundaries found in this study and numbered with gray boxed numerals. Bold Roman numbers with or without subscripts are the names of hydrogen phases. Other symbols and lines are indicated in the figure legend. Two green dashed lines indicate the position of frequency-temperature slope change within phase $I_{hb}$ and $I'$ respectively; two thicker green solid lines overlapped with phase lines 8 and 10 are the boundary of starting pressure for strong anharmonic structure transitions (see text). Light blue lines overlapped with phase line 9, 10, 11, and 12 are the boundary of starting pressure for the appearances of new peak $v_1'$.

Their locations clearly suggest phase boundaries. Points found from the isothermal plots are mostly located on those boundaries approximately in the vertical direction, and the points found from isobaric plots are mostly located in the horizontal boundaries. Curved and diagonal boundaries are constructed by the mixed data points found from both the isothermal and isobaric plots. This is critical evidence showing that the results from two sets of thermodynamic mapping are self-consistent. A total of 13 boundaries can be constructed with 15 possible phases of solid hydrogen existing in the PT range studied.

## Observations and interpretation

The shapes of the FP dependence for vibron $v_1$, at all temperatures display a great similarity in the two pressure regions, as shown in the selected isothermal plots of Figure 5. The phases inside of the first region are confined by phase boundary 5 and melting line below ~150 GPa, and the phases in the second range below ~250 GPa is confined by phase lines 5, 8, 10 and the melting line as shown in Figure 4. In these two pressure regions, only one vibron was observed. The phase transitions were characterized mainly by discrete shifting in both vibron frequency and peak width. Although the observed discontinuities of vibron frequency in these two regions are quite obvious, the changing magnitudes are generally minor in some places. The increase of peak width as function of pressure and temperature does not show a noticeable discontinuity in the defined phase lines except for phase line 5 which is the boundary between these two areas. We postulate that these features of vibron $\nu_1$ are an indication of structure similarity in these two phase areas; most likely representing some kinds of iso-structural transitions. Therefore, the new phases found in these two regions are labeled on the basis of their low temperature identification label with subscripts "h" and sequential alphabetical letters meaning high temperature phases. Four new phases at positions above phase I in the first phase region are named as $I_{ha}$, $I_{hb}$, $I_{hc}$, $I_{hd}$, whereas two new phases at positions above phase I' in the second region are named as $I'_{ha}$, and $I'_{hb}$. This typical assignment helps remove ambiguity and confusion with previous studies.

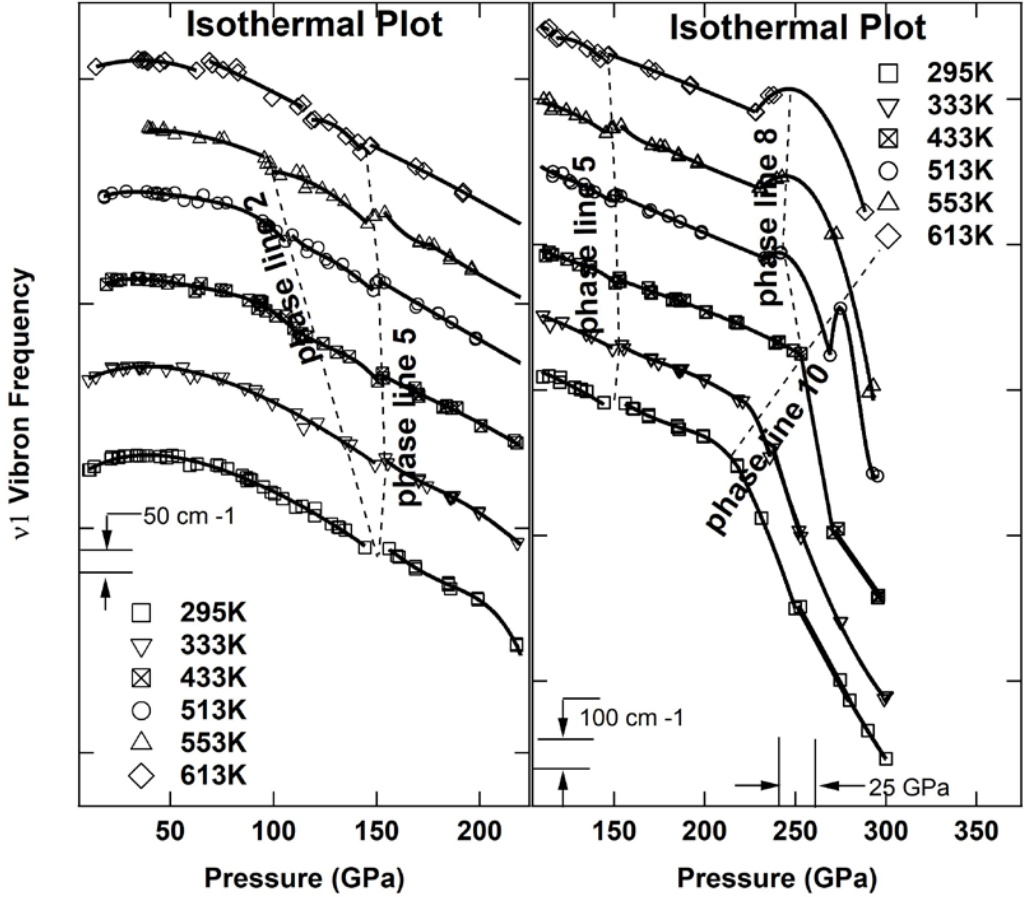

Figure 5. Isothermal FP dependence of $\nu_1$ for selected temperatures show similar shapes before the downward slopes happening at phase line 10 and 8 (see figure 4 for the two line locations). For clarification, vertical axes in both panels are offset. At even higher pressure range, the structures of FP dependence show multiple changes reflecting complex phase transitions in the areas (right panel). The dashed lines connecting some obvious FP discontinuous points are some of corresponding phase lines as mentioned in text.

In the first pressure region, previous studies performed below or at room temperature suggest that phase I transformed to phase I' at ~143 to ~150 GPa at 300 K,[19,20] seemingly consistent with the critical point formed by phase lines 2 and 5 of this study at 149 GPa and 295 K (Figure 4). Therefore, a new phase $I_{hb}$, to the best of our knowledge, which has never been predicted, must exist between I and I' at temperatures above 300 K. We also found that the discontinuities of frequency-temperature/pressure dependencies along with these two-phase lines decrease with decreasing temperature and reach almost undetectable values (see Figure S2 #15 - #20 and Figure S1 295 K and above in SM) at room temperature respectively at this point. This might be the reason why this discontinuity has not been observed by many previous isothermal compression studies at room temperature. An interesting finding is that if we systematically check the temperature dependence of vibron frequency from Figure S2 of SM, the temperature slope for isobaric vibron frequency corresponding to the new phase $I_{hb}$ area changes from a positive value at lower pressure to a negative one at higher pressures. The turning point appears at pressures around ~120 – ~128 GPa (see Figure S2 #15-16 of SM). However, no sharp discontinuity was found in isothermal FP plots (see Figure S1 of SM) crossing this area, indicative of possible existence of a higher-order transition boundary as shown as a green dashed line in Figure 4.

The FP dependence of vibron $\nu_1$ in the second pressure range is a group of lines having similar trend shape but with very small frequency shift along phase line 5. This similarity ends at phase line 10 or phase line 8 (see Figures 4 and 5). There are two phase lines, 6 and 7, found in this area based on isobaric or isothermal frequency discontinuity of vibron $\nu_1$, which forms phase $I'_{ha}$ and $I'_{hb}$ located in the higher temperature area above phase I'. Similar to phase $I_{hb}$ in the first pressure range, phase I' shows a ridge where the slope value of the FP dependence turns the sign from negative to positive at ~195 GPa (see Figure S2 of SM before and after #24) but has no sharp discrete frequency shift in isotherms (see Figure S1 of SM for temperature below 473 K). We speculate that another possible higher-order transition possibly exists, as indicated by a dashed green line in Figure 4.

Note that, large Raman lattice mode variation was found in the pressure range between ~216 and 250 GPa at room temperature (Figure 7 panel **a**); on the other hand, the phonon peak $L_3$ is presented at all temperatures below 250 GPa (Figure 7 panel **b** and **c**). This observation may imply that some kind of structural similarity is continuing in the range during the phase evolutions.[21-24]

Much more complicated phase transition pictures are found when pressure-temperature reaches over phase lines 10 and 8. Transition around ~220 ± 15 GPa and 300 K was reported previously,[9,10,13,15] featuring a sudden change in the FP slope of $\nu_1$ and peak width broadening. Also, a weak second vibron $\nu_2$ as well as a lattice mode $L_1$ have started to appear. Theoretical simulations indicate that the phonon frequencies of a new layered structure are consistent with the experimental observations (phase IV).[4,25-29] This transition is consistent with our finding in the present study at 216 GPa and 295 K on phase line 10. However, there are disputes concerning the $\nu_1$ vibron behavior after this point that defines the phase boundary between phase III and IV (Figure 4). Previous experiments all imply I – III – IV phase sequence at 300 K. It means that, after the transition mentioned above, hydrogen enters a narrow phase III range before transformation to phase IV. However, the reported pressure range for phase III along the isotherm of 300 K was inconsistent among different studies.[14,30,31] The phase diagram proposed by Eremets et al. gives a range approximately between ~204 to ~216 GPa,[31] whereas Howie et al. find it between 190 ~ 235 GPa based on the spectroscopic measurements.[10] In the two reports cited above, only the upper bounds of these pressure ranges are associated with the appearance of the second vibron peak ($\nu_2$). If this is true, it implies that the sudden FP slope downward trend and the peak width broadening of $\nu_1$ are separated from the appearance of the second vibron peak $\nu_2$ in the proposed layered structure. This is inconsistent with theoretical predictions.[26,29] Our finding suggests this pressure range should be between 216 and 250 GPa. The second Raman vibron peak of $\nu_2$ as well as the lattice mode $L_1$ appear simultaneously with the down slope of $\nu_1$ frequency and peak width broadening at 216 GPa and 295 K though the intensities of the former are very weak in the phase III range.[15] It is important to note that the reported room temperature slope of FP dependencies for vibron $\nu_1$ after entering phase III are consistent between different groups (see Figure S3 of SM), but very different from that observed at low temperature of phase III.[32-34] Combined with the appearance of weak peaks of $\nu_2$ and $L_1$, the structure of solid hydrogen in the phase III region might not be simple. Possibly some modified structures exist at different temperature portions, and more detailed PT mapping work needs to be conducted in the phase III region.

The room temperature transition point for phases III - IV at 250 GPa displays anomalous property changes reported previously[15]. Here, new evidence that can be added to the transition is a slight kink in the FP dependence for vibron $\nu_1$ as shown in the isothermal plot (Figure S1 of SM at 295 K). Although the small discontinuity can be easily missed at single isothermal compression experiments conducted previously,[9,10,15,35] the fact that it is located right on the III – IV phase boundary (phase line $IR_2$ in Figure 4) determined by previous infrared (IR) experiment apparently cannot be discounted as an accident.[14] In addition, two other discontinuity points at 333 K while the pressure is at 235 (Fig. S2 of SM, #26) and 236 GPa (Fig. S1 of SM, #3) respectively, determined from both isobaric and isothermal plots of this study, also are located on the infrared-determined boundary $IR_2$ (as shown in Figure 4).

A weak second vibron peak $\nu_2$ at ~4200 cm$^{-1}$ found experimentally[9,10,15,31,35] at ~ 216 - ~235 GPa and 300 K has stimulated intense theoretical work in recent years. It has been widely attributed to a layered structural change.[4,25-29,36] Causatively, the disappearance of this high frequency vibron of $\nu_2$ would be the indication of a phase change (or decomposition) for the layered structure. In this study, the layered structure was found unstable at high temperatures as indicated by the disappearance of vibron $\nu_2$ during heating. Figure 6 shows the spectra of solid hydrogen in five heating processes started at different pressures. The $\nu_2$ vibrons in all five runs weakened upon increasing temperature and disappeared at various temperatures in different runs. The PT points for the last appearance of $\nu_2$ peak during heating are plotted in Figure 4, they nicely coincided with phase line 10 and infrared-determined $IR_1$.[14] It seems that the line can be extended smoothly to align with the I' – III transition line at low temperature. Phase line 10 represents a significant boundary which divides the phase space of solid hydrogen into two portions, distinguished by presence and absence of a second vibron on each side. A more detailed picture is that the appearance of a significant downward turn for FP slope and the $\nu_2$ peak disappearance happens simultaneously at the initial stage and proceed together for a while. The temperature values for this situation increase with pressure along phase line 10 from ~216 GPa at 295 K to ~253 GPa and 433 K, which is the intersection of phase lines 10 and 8 in Figure 4. At this point, pressure for the downward turn of FP slope stops increasing, and the downward slope appears along phase line 8 while vibron peak $\nu_2$ disappearance is still tracking with phase line 10 to the highest pressure of this study (see Fig. S1 of SM). Discontinuous frequency points of vibron $\nu_1$ detected through isothermal and isobaric plots are also located along phase line 10. This indicates that the phase transitions crossing over this phase boundary show both mode appear-disappear and frequency discrete shifting. The consistency between two types of measurements confirmed reliability of line 10 as a phase transition boundary. Collection of all the information, allows us to conclude that phase lines 10 and IR$_2$ represent the higher and lower temperature boundaries of phase IV, respectively (Figure 4).

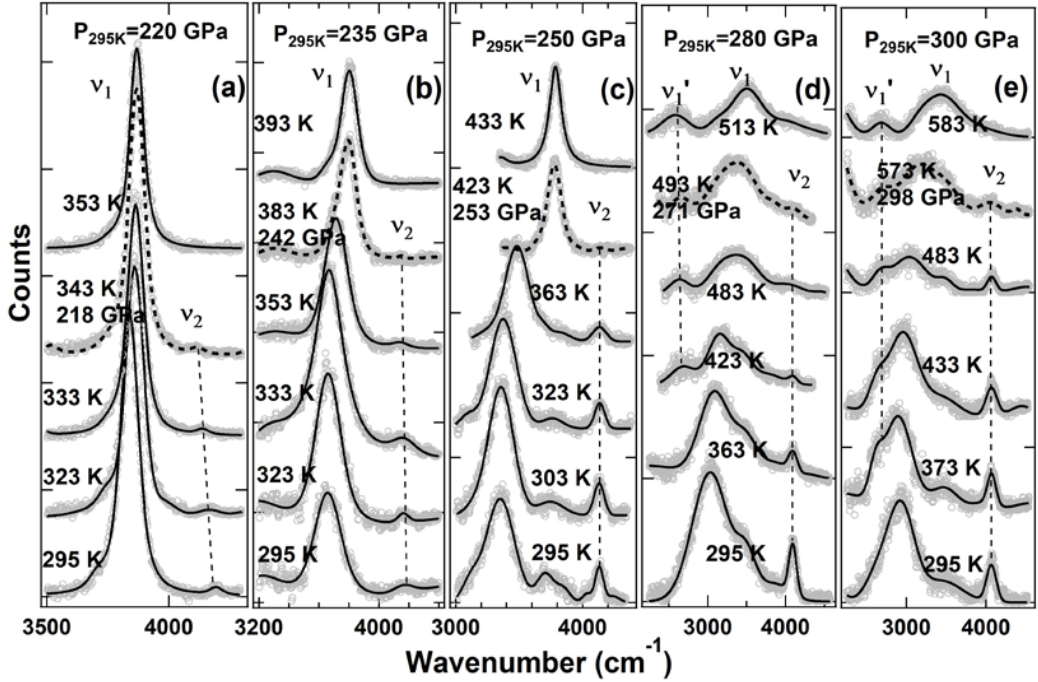

Figure 6. Vibron $\nu_2$ spectra change with temperature during the constant force-loading heating at different starting pressures. The dotted spectrum at each panel shows the last appearance of vibron $\nu_2$ at this heating run. A new peak $\nu_1$' appears at two highest pressure heating runs as shown in panel (d) and (e). Unmarked peaks (or shoulders) to the right of $\nu_1$ are the fluorescence peak (see figure S4 of SM), its intensity seems diminished with increasing temperature.

We wish to point out the existence of an important phenomenon for vibron $\nu_1$ revealed in this study. It is apparent that the significant downward slopes of FP always come along with peak widths that suddenly broaden at all PT conditions (as shown in Figure S1 and its insets of SM). This is an indication of structure anharmonicity. The PT trend along with phase line 10 and 8 discussed above is shown as a thicker green solid line in figure 4. High pressure phases to the right side of this green solid line are all involved. This finding could inspire the future theoretical studies on structural investigation of the hydrogen system.

The layer structured hydrogen shows even more complicated phase transitions at the higher pressure-temperature regions. Previous room temperature studies found a very slight pressure slope change of band gap and Raman vibron $\nu_1$.[30] Also, an abrupt intensity change for lowest libron $L_1$ around 275 – 285 GPa was reported.[15] These have been attributed to a structural transition to phase IV'. The reported observation for loss of the Raman mode and the sample turned to a dark, electrically conducting state at similar pressure was not reproduced.[9] The phase line 11 found in this study reveals that FP slope changes for vibron $\nu_1$ around this pressure range are very weak at temperatures below

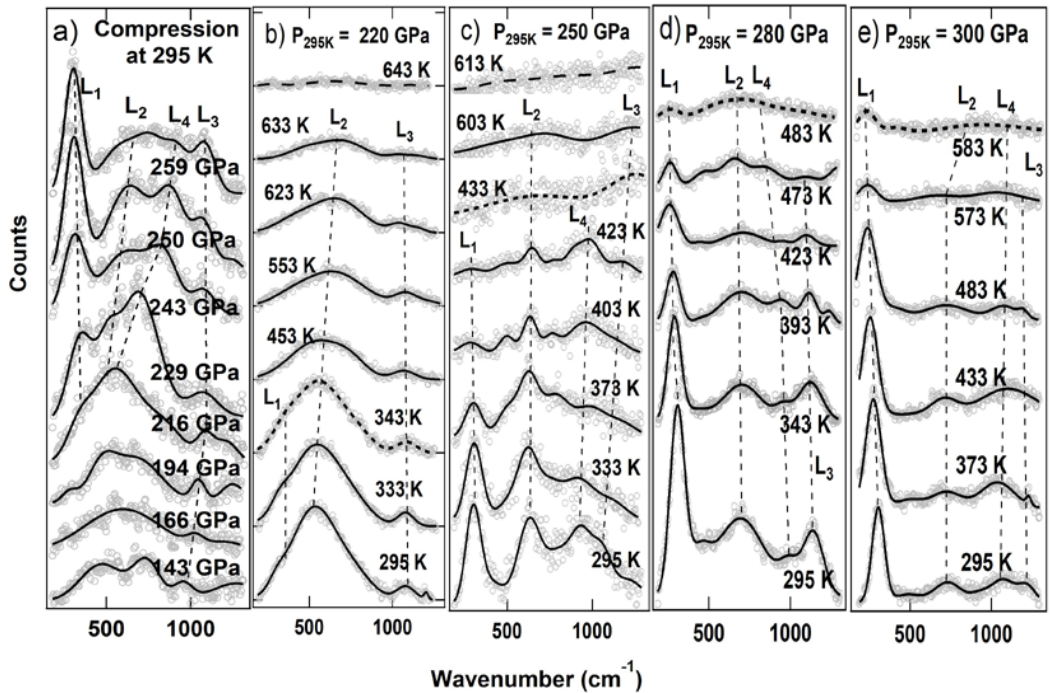

Figure 7. The behavior of librons at high pressures and temperatures. Panel a) shows the libron spectra under isothermal compression at 295 K. The broad libron peak at lower pressure is split into $L_1$, $L_2$, and $L_4$ after 216 ~ 220 GPa which is the I' to III phase transition point at room temperature. Panels b) to e) show how the librons change with increasing temperature at selected heating runs. Dotted line spectra are collected at the PT points corresponding to $\nu_2$ vibron disappearance (see figure 6). Dashed line spectra in panel b) and c) are collected at the melting point. In the heating runs with starting pressure ≤ 250 GPa in panels b) and c), $L_1$ and $L_4$ (the $L_4$ can be seen only in panel c) because it has not separated enough from $L_2$ at 220 GPa) disappear simultaneously along with disappearance of $\nu_2$ vibron, while $L_3$ does not and persists to the melting line; these happen within phases I', I'$_{hb}$, and IV$_{ha}$. However, heating runs with starting pressure of 280 and 300 GPa in panels d) and e), both libron $L_1$ and $L_4$ no longer disappear simultaneously with $\nu_2$, instead, only $L_3$ does; these happen within phase VI$_{ha}$.

353 K (see Figures 4 and S1 of SM). Instead, the intensity change of lowest lattice mode $L_1$ in this pressure region is very obvious even below 353 K as shown in figure 7 (compare panels **c** and **d**). However, FP slope change for vibron $\nu_1$ becomes very obvious starting from 353 K up to 513 K (Fig. S1 of SM). This feature characterizes the higher temperature portion of phase line 11. Obviously, phase line 11 is the boundary between phase IV and IV'. The PT point of 277 GPa and 513 K is the intersection point of phase line 11 and 10 (Figure 4). At temperatures above this point, $\nu_2$ disappears (see Figure 6 panel **d**), indicative of the existence of a new phase. The slope of FP dependence for $\nu_1$ in this region drops down steeply similar to that in phase IV (see pressure range higher than phase line 8 in Figure 5), but very different from phase I'$_{hb}$ (which is the only region with positive slope for FP dependence), so it is named as phase IV$_{ha}$.

Room temperature studies have also reported another possible transition to phase V around ~300 to 325 GPa at 295 K[15,35,37] or vicinity temperature range.[37] Here, we found that phase IV' seems to be terminated at phase line 13 as defined from the slight frequency discontinuity of $\nu_1$ in isothermal plots at 333 K and 353 K. It is a little lower than 300 GPa (Figure S1 of SM), consistent with the previous report. This short line ends when temperature increases to 373 K, at which point a new peak $\nu_1$' starts to appear (see Figure 6 panel **e**). This indicates that another new phase transition happens at a higher temperature region. Pressure higher than phase line 13 should be the phase V area (see Figure 4). No additional data are available to define the phase boundaries in higher PT range for phase V.

As already mentioned above, an interesting thing to note is that a new peak $\nu_1$' is found in two higher pressure heating runs. During the first heating run with starting pressure of 280 GPa, the peak appears when temperature reaches to 423 K and the *in-situ* pressure at 273 GPa. The peak is on the lower frequency side of vibron $\nu_1$ (panel **d** of figure 6) at ~2690 cm$^{-1}$. The same peak appears also in the heating run at a starting pressure of 300 GPa, but a lower temperature of 373 K (see Figure 6 panel **e**) and *in-situ* pressure of 299 GPa with a similar frequency. These have been attributed to the vibron mode of new phase *Cmca-4* reported previously.[17] Because of this new peak, a new phase boundary can be drawn from these two PT points as phase line 12, together with phase line 10 and 11 form a new phase VI as illustrated in Figure 4. This phase is featured with three Raman vibron modes of $\nu_1$, $\nu_2$, and $\nu_1$'. Figure 8 shows temperature-frequency trajectory during these two heating runs. Frequency of the $\nu_1$' shows an approximate stable pressure dependence when it appears, but a slightly different temperature dependence during these two heating runs.

Upon further heating from the phase VI to cross over phase line 10, phase VI transforms to another new phase because it lost vibron $\nu_2$, with only two vibrons $\nu_1$ and $\nu_1$' remaining. The heating path started at 280 GPa is along with phase line 9 and thus enters the region of phase IV$_{ha}$ (see Figure 4). As mentioned above, phase IV$_{ha}$ has vibron peak $\nu_1$ only, but there are currently two vibron peaks. Obviously, phase line 9 is a new phase boundary as indicated by the appearance of the new peak. The name for this new phase is VI$_{ha}$. It is unusual that phase line 9 does not show a frequency discontinuity for vibron $\nu_1$ when checking the FP dependence between these two phases from the isothermal plots (see #13 to #17 in Figure S1 of SM). Pressure-induced phase transition involving the appearance of new peaks usually is accompanied by a discrete shift of the peak frequencies. The continuous $\nu_1$ frequency from phase IV$_{ha}$ to VI$_{ha}$ implies that new peak

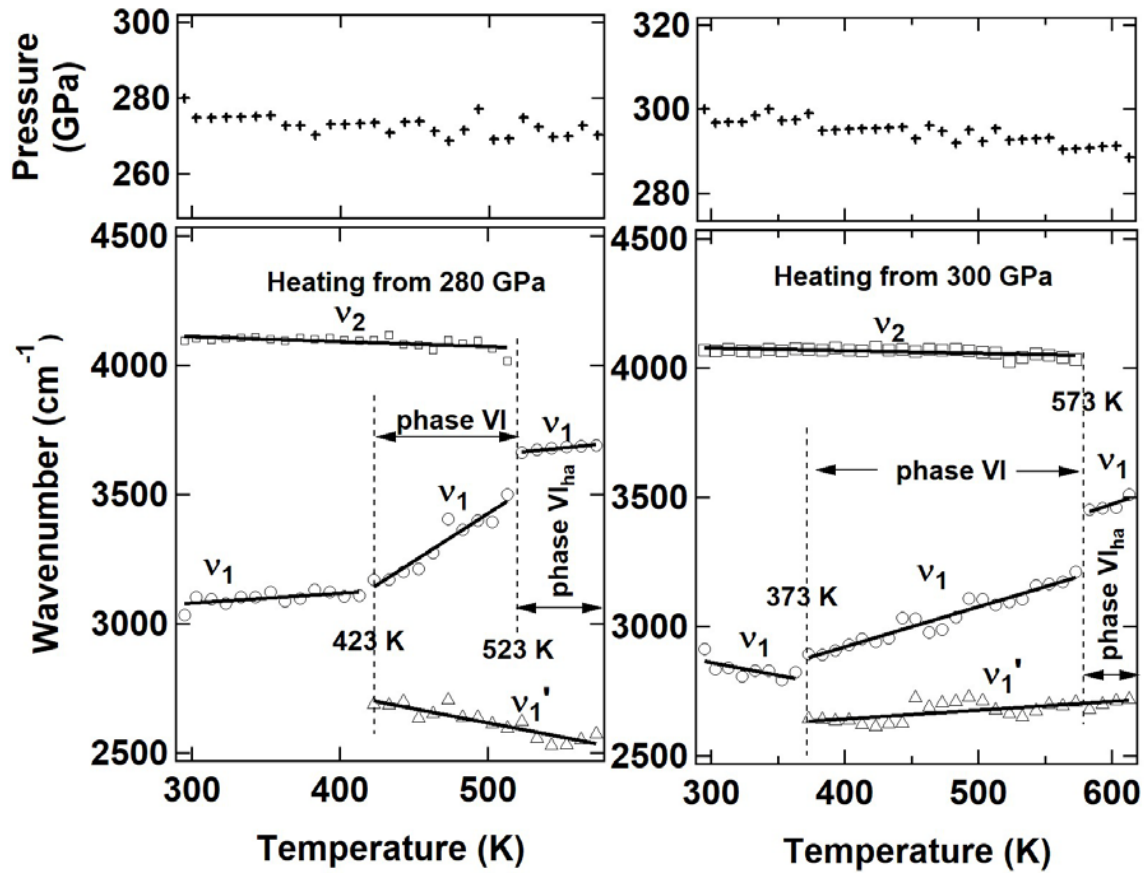

Figure 8. Temperature dependence for three vibration modes at two heating runs revealed interesting structure relations between the possible phase transitions during the heating. Symbols: measured data, solid lines are for eye guiding. Double arrows show temperature range for phase VI and VI$_{ha}$ during two heating runs. Frequency modulations by pressure variation during heating were not made because no reliable isothermal fitting data are available at this high pressure range.

$\nu_1$' does not have a structural relation with vibron $\nu_1$ in phase VI$_{ha}$. Therefore phase VI$_{ha}$ may not be a new single phase with two vibrons. Phase VI$_{ha}$ most likely is a multiple phase, where $\nu_1$ and $\nu_1$' belong to different phase components. On the other hand, the continuous $\nu_1$ does not necessarily mean the structure, which $\nu_1$ belongs to at phase area VI$_{ha}$, is the same as that of phase IV$_{ha}$. It might only be just a numerical coincidence for the frequency. This assertion is supported by the lattice mode behavior at high temperatures as shown in Figure 7. The evidence presented in panels **d** and **e** of Figure 7 clearly shows the lattice mode behaviors of phase VI$_{ha}$ are different from that of phase IV$_{ha}$ (panel **c** of Fig. 7). Librons $L_1$ and $L_4$ along with $\nu_2$ vibron disappeared at 433 K, but not for $L_3$ in phase IV$_{ha}$. However, heating runs in panels **d** and **e** of Figure 7, show that both libron $L_1$ and $L_4$ no longer disappeared with $\nu_2$ at 483 or 583 K. Only $L_3$ disappeared along with it, this happened in the phase VI$_{ha}$. The disappearance of lattice mode $L_3$ within phase VI$_{ha}$ indicates that phase component contributing $\nu_1$ vibron likely no longer belongs to hcp structure system according to the analysis of previous studies,[21,23,24] and is apparently different from phase IV$_{ha}$

More structural information represented by the behavior of peak $\nu_1$, $\nu_2$, and $\nu_1'$ during heating may be extracted from Figure 8. First, the appearance of $\nu_1'$ and disappearance of $\nu_2$ is always accompanied with a discontinuous frequency change in $\nu_1$. This means that $\nu_1$ is connected with both $\nu_2$ and $\nu_1'$. Second, when $\nu_1'$ appears, or $\nu_2$ disappears, nothing happens on the other one, indicating no connection between these two modes, i.e. new peak $\nu_1'$ is only related with the graphene-like layer of phase IV' (assuming the structure of phase IV' is very similar to phase IV as previously proposed[30]). The above observations suggest the $\nu_1'$ might originate from a progressive dissociation of graphene-like layers at higher temperature. The dissociation would result in a frequency discontinuity of $\nu_1$ whereas disappearance of $\nu_2$ might result in the second large shift in $\nu_1$ but not in $\nu_1'$. After dissociation, $\nu_1$ and $\nu_2$ are still from two layers of a phase component, which is the phase non-related to $\nu_1'$. The component from dissociation featured with peak $\nu_1'$ is a separated entity, co-exists before and after the disappearance of $\nu_2$, indicating that both VI and VI$_{ha}$ exist in multiple phase areas (see Fig.4 and 8). We note that, although the frequency uncertainty due to pressure change during heating in Figure 8 was not corrected, we think the total change of ~10 GPa for the entire heating run, without obvious sudden jump up or down was recorded, does not affect the main frequency feature discussed here.

On the other hand, the disappearance of $\nu_2$ at higher temperatures indicates that the structure for the disordered molecular layer of the phases (presumably includes phases III, IV, and VI) is unstable upon heating; it dissociates at temperatures above phase line 10, and reforms into the new phases of I', IV$_{ha}$, and VI$_{ha}$. The transition is also accompanied with $\nu_1$ frequency discrete shift from phases III, IV, and VI. As discussed above, since both the phase VI and VI$_{ha}$ regions are probably the co-existing multiple phases, we can draw light-blue lines in Figure 4 as the boundary of multiple phase regions in which $\nu_1'$ exists. The region is enclosed by phase lines 9, 10, 11, and 12 in solid phase area with open side to the liquid area. Although not enough data are used to ascertain the behavior of $\nu_1'$, we cannot rule out the possibility for the existence of peak $\nu_1'$ in the liquid area as seen in the Figure 2 **d** and **h** of Ref. 17.

Figure 8 shows that $\nu_1$ frequency decreases significantly with increasing pressure, but $\nu_1'$ is relatively stable. The frequency difference between $\nu_1$ and $\nu_1'$ therefore decreases with increasing pressure, and the two peaks are expected to overlap at higher pressures. If $\nu_1'$ owes its origin to the progressive dissociation of the graphene-like layer of phase IV' as suggested above, the overlapping should indeed be a replacement. It means that the $\nu_1$ peak should finally be replaced by $\nu_1'$. Further experiments at higher pressure are required to confirm this prediction.

The appearance and disappearance of the multiple vibron peaks in the region bordered by light-blue lines offers rich information for understanding the phase evaluation including the metallization of solid hydrogen. Crystal structure prediction based on density-function theory (DFT) and *ab initio* random structure searching algorithm have been mostly used towards predicting possible structure candidates of solid hydrogen at high pressures.[3,4,29,36,38-41] Phonon frequency spectra of candidate structures calculated over a wide pressure range for comparing with experimental results were the direct way for checking the reliability of the theory. Among many candidate structures, phonon frequencies for *Cmca-4*,[17,41] *I4₁/amd*, and *R-3m* structures[38] are reported to be close to that of vibron $\nu_1$' found here though the predicted stable pressure ranges are different. A common character of these candidate structures is that they are all metallic. Our phase mapping has revealed spectroscopic evidence that phonon frequency consistent with most possible metallic states indeed appears at this particular region of hydrogen phase diagram. The metallic transition process may not simply have a sharp, single boundary. It could be a multiple transition process. The phonon frequency $\nu_1$', which serves as a sign of metallic transition, prefers to appear at higher temperature while pressure could be lower or vice versa, consistent with the big picture built by most previous reported experiments[37,42-46] and theoretical predictions.[3,4,39,47-50]

## Summary

This study presents a large amount of physical information for dense solid hydrogen in a broad pressure-temperature range, which occupies an important phase space not known previously. The data analyzed here are collected during many independent constant force-loading heating runs, and each of them includes numerous stepwise temperature increments and different starting pressure respectively. The collected experimental raw data are subjected to hybridization effects of both pressure and temperature changes unavoidably introduced by heating. The commonly used "reversal" method, which consists of repeatedly changing temperature back and forth while keeping pressure constant for in situ verifying any possible transitions, therefore is not practical. A network data-converting method has to be used for constructing the isothermal and isobaric dependencies of the properties of Raman modes to determine phase transition boundaries. The results demonstrate that not only most previously identified transitions reported by room temperature compression experiments are nicely verified; they also show the sensitivity for finding some minor changes which could be the key for resolving controversial disagreements remaining from previous studies. We believe that the finding of transitions located at lower pressure but higher temperature range in this study, manifested by minor spectroscopic changes and not predicted previously, also benefits from this sensitive capability. Those small changes may have unknown significance in the future study for this important prototype material in this virgin phase territory, although we agree that more future research for verification of those minor transitions is necessary. This is a logical accomplishment resulting from the reduction of some of the systematic uncertainties; that objectivity is obtainable only by collecting a large amount of data at random PT points. The most interesting findings in this study might be those located in the PT range greater than ~220 GPa and high temperature up to the melting line. Molecular hydrogen likely undergoes a gradual dissociation process before its metallization transition. Explanations of such a phase evolution will be of great theoretical interest in the future. It certainly will revise our current theoretical understanding of this simplest element.

## Methods

Technically, heating hydrogen samples under extremely high pressure is a highly challenging operation; obtaining precision scientific data at these conditions would be a severe trial. We have used a special symmetric externally heated diamond anvil cell (DAC) equipped with double compact resistive heaters, closely surrounding the anvil-gasket area from both sides. This not only ensures a uniform temperature distribution across the sample, also offers very high heating effectiveness such that a short heating duration is available; that reduces the chance for chemical reactions between hydrogen and other parts. Different gasket materials were used in this study for checking possible chemical reaction between gasket and hydrogen.[17] There is no detectable evidence showing the existence of possible hydrogen-diamond reaction products as reported recently.[51] Thermocouple feed-back electronic system can offer a constant temperature period between each incremental change for conducting spectroscopic measurements which characterize the sample, diamond anvil, and background. Special design considerations are needed in order to keep heating area distantly and thermally well isolated from the mechanical tightening parts of the DAC. This is critical for maintaining pressure-temperature stability. A sample diameter of 15 microns may be obtainable for pressure up to 250 GPa, and ~5 to 7 microns is the typical size of sample for reaching 300 GPa. An in-house designed con-focal optical Raman spectroscopic system with fiber optics was used for sampling an area of only few microns of sample. Extremely careful system alignment and focusing conditions are necessary for obtaining highest quality, and most reliable spectroscopic data at these extreme conditions. For the pressure measurement, one or two tiny ruby or $SrB_4O_7:Sm^{2+}$ chips[52,53] can be used for monitoring pressure during heating at lower pressure/temperature range, but all higher pressure experiments are conducted without any pressure sensor in the gasket hole. Instead, first-order diamond Raman from the anvil tip contacting the sample was measured and monitored during the heating. The details of special design and operation techniques for heating DAC are published in the supplementary materials of Ref. 17.

*In situ* pressure and temperature measurements are obviously critical to provide precise and accurate data for constructing a phase diagram. Pressure measurements for most portions of the diagram use diamond Raman-pressure relation method. There are many studies on the calibration of pressure dependence from the diamond Raman signal establishing it as a practical pressure scale. However significant discrepancies have been reported. The reason for these discrepancies is complicated as the anvils are under extremely high uniaxial stress which changes with crystal orientation and geometrical configuration of polished anvil tips, gasket materials, and the strength/hardness of different samples etc.. The complexity of Raman signals under those conditions may be responsible for the discrepancies. Since the recent proposed pressure scale by Akahama et al.[54] was used by different reported hydrogen studies,[15,31,35,37] the pressure dependency of the measured vibron frequencies at room temperature are in reasonable agreement with each other below 300 GPa with an estimated disagreement of ~ ±3% from the average of the measurements (see Fig. S3 of SM). A practical method to determine the pressure at simultaneously high P and T in this study is based on two portions of pressure determination. The starting pressure at room temperature is always checked by the averaged pressure dependence of vibron frequency obtained previously,[15] and it is then corrected with the variation during heating for obtaining the pressures at high T. The pressure variations during heating are the difference calculated from Akahama's scale, by inputting diamond Raman shift recorded at room temperature and each temperature increment respectively. Frequency-temperature shift at room pressure for diamond Raman is taken into account with assumpion of no pressure cross-effect for its thermal shift.[55] This method was validated by in situ high P - T x-ray diffraction and Raman scattering at synchrotron radiation facility as mentioned in the supplementary materials of Ref. 17. Temperature measurement was determined by the straightforward method employing thermocouples placed close to the tip portion of diamond anvils. We find that the temperature gradient between gasket hole and the thermocouple contacting point should be the main uncertainty for temperature measurement, testing experiment indicates a 13 degree gradient could exist between these two places when sample reaches 730 K (~1.8%).

Also, vibron spectra are strongly affected by fluorescence at pressures above 250 GPa (see Figure 6). Two-wavelength Raman-excitation method has to be used for distinguishing it, as shown in the Figure S4 of SM example.

## Acknowledgement

The authors are grateful to Z. Gebarlle, T. Strobel, John Tse, R. J. Hemley, T. S. Duffy, and Y. M. Ma for useful discussions. The original data re-analyzed in this work were collected under the support of CDAC (Carnegie/Department of Energy Alliance Center, Grant No. DE-NA-0002006) and EFree (Energy Frontier Research Center, Grant No. DE SC0001057) research programs. CZ is grateful for the Carnegie community invaluable support. The CHESS is partially supported by the NSF awards of DMR-1332208 and DMR-1829070.

**\*** Author to whom correspondence should be addressed. czha@carnegiescience.edu

# Supplementary Materials
# Experimental phase transition mapping for hydrogen above 300 K up to 300 GPa

**Chang-Sheng Zha[1]\*, Hanyu Liu[2], Zhongwu Wang[3], William A. Bassett[4]**

[1]Earth and Planets Laboratory, Carnegie Institution of Washington, 5241 Broad Branch Rd. N.W., Washington DC, 20015, USA
[2]International Center for Computational Method & Software, and State Key Lab of Superhard Materials, College of Physics, Jilin University, P. R. China
[3]Cornell High Energy Synchrotron Source (CHESS), Cornell University, Ithaca, New York 14853, USA
[4]Department of Geological Sciences, Cornell University, Ithaca, New York 14853, USA

Figure S1. Isothermal frequency and peak width plot of hydrogen vibron $\nu_1$, symbols and solid lines are experimental data and fitting lines respectively. The small arrows point to frequency discontinuities and accompanies with pressure values. Boxed number indicates the phase line of figure 4, in which this P T point located. Gray Roman letters are the phase name of the indicated range.

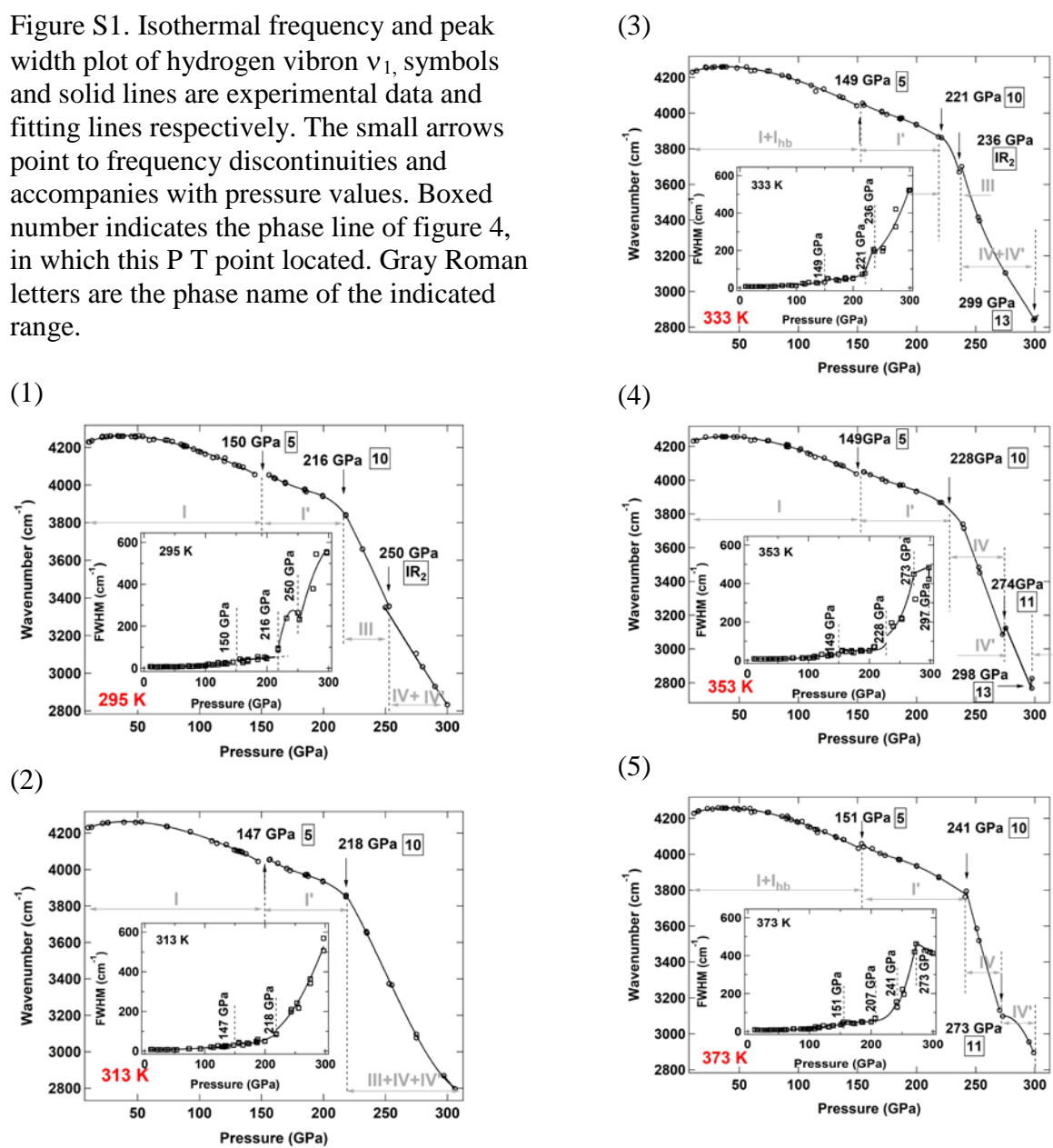

(6)

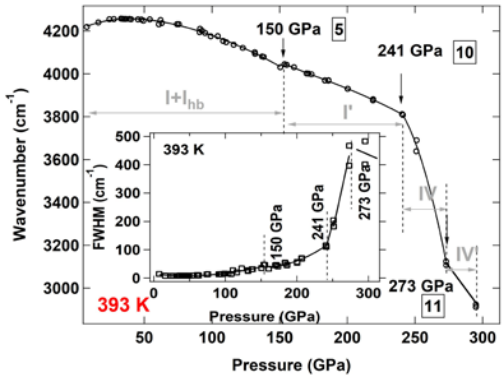

(7)

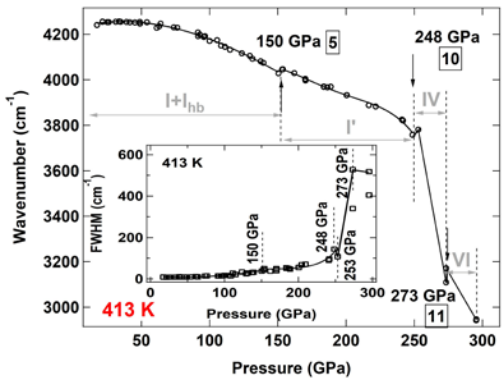

(8)

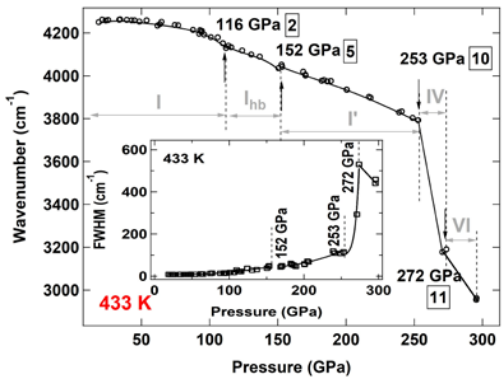

(9)

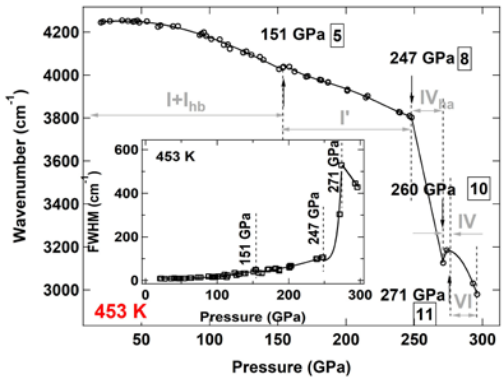

(10)

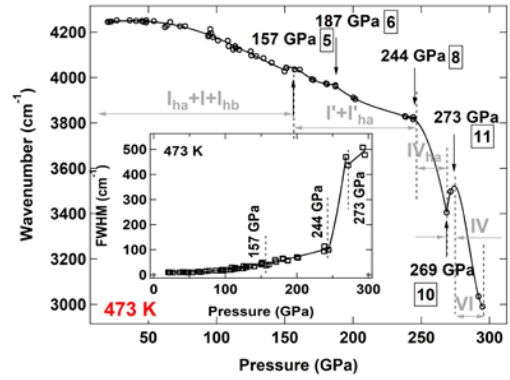

(11)

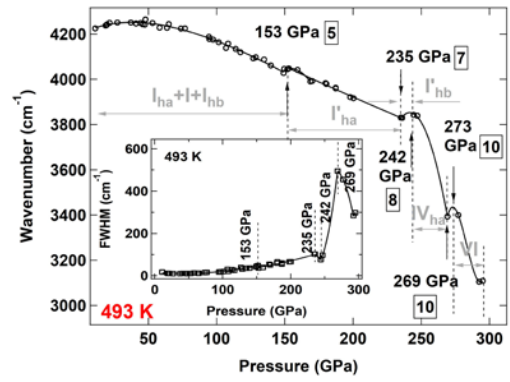

(12)

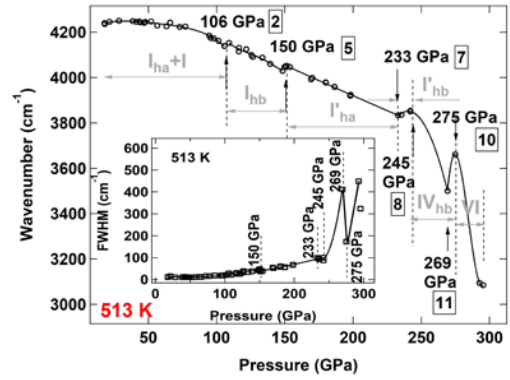

(13)

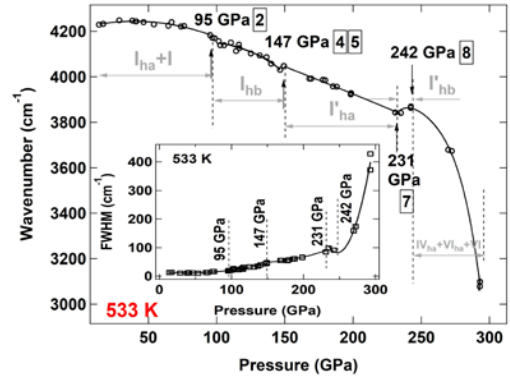

(14) 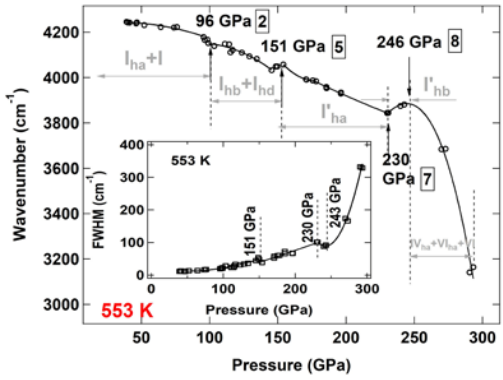

(18) 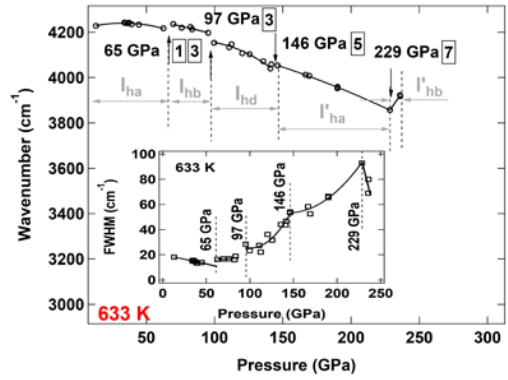

(15) 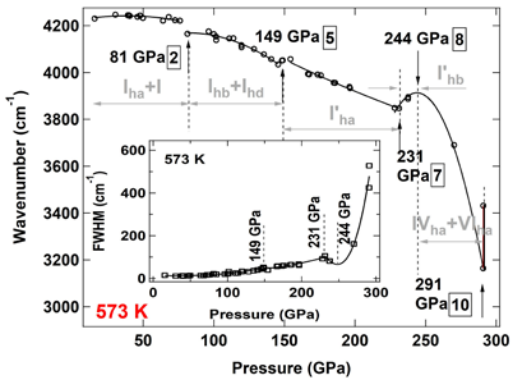

(19) 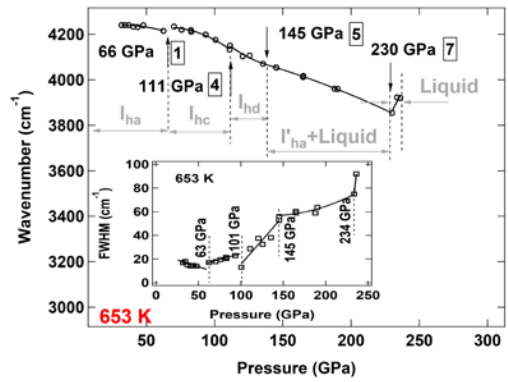

(16) 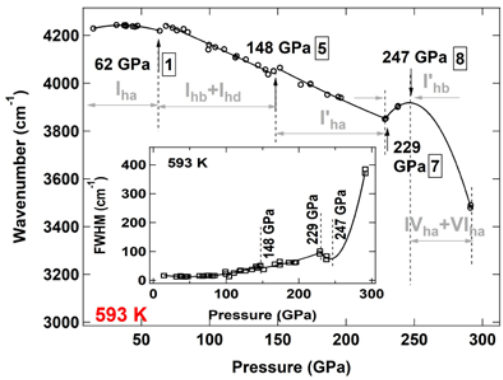

(20) 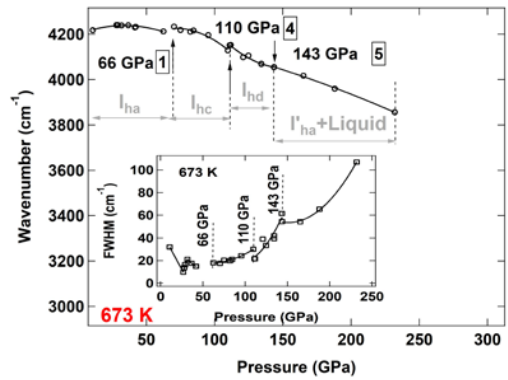

(17) 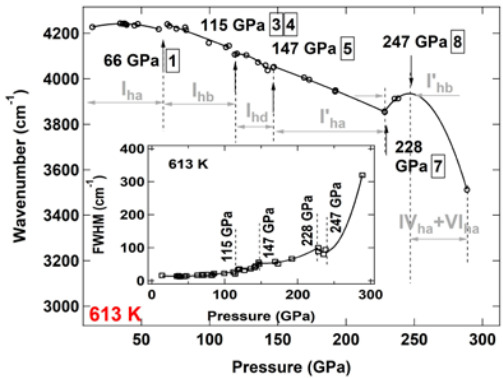

(21) 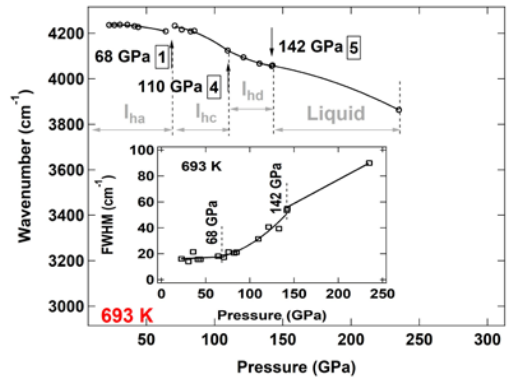

(22)
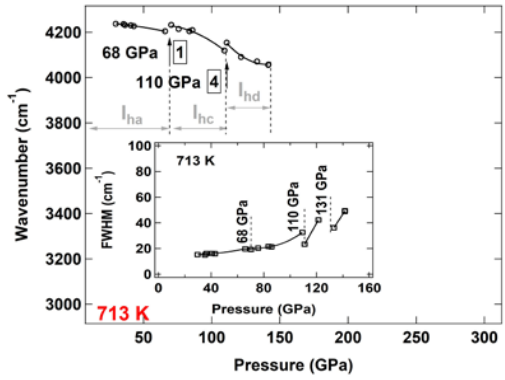

(26)
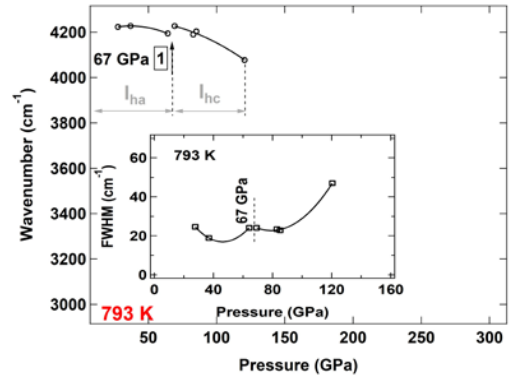

(23)
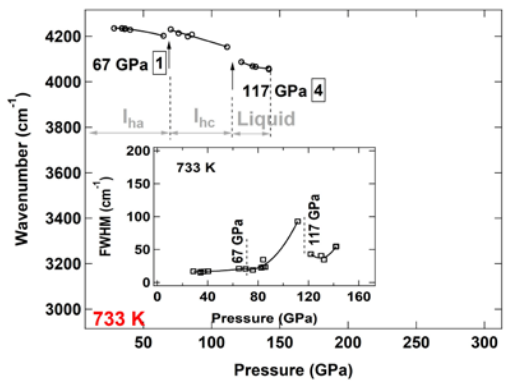

(27)
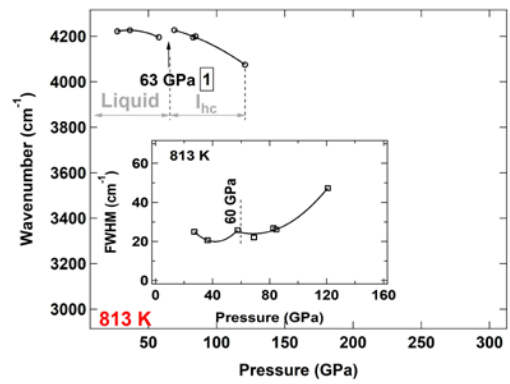

(24)
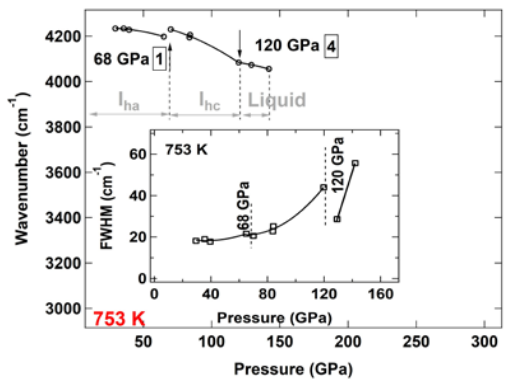

(25)
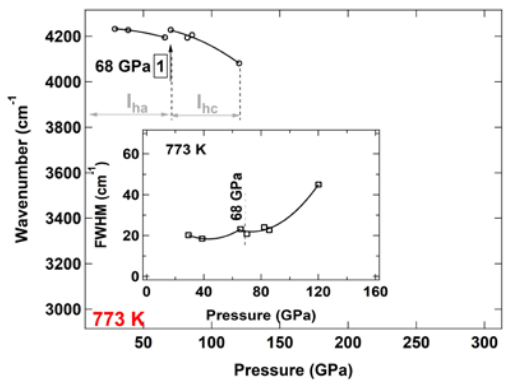

Figure S2. Isobaric frequency and peak width plot for hydrogen vibron $\nu_1$. Open symbols are experimental data, solid circle is isobaric data calculated from fitted isothermal frequency-pressure dependences. Solid lines are for eye-guiding. The boxed number indicates the phase line of figure 4 at which this discontinuity point located. Gray Roman letters are the phase name of the indicated range.

(1)

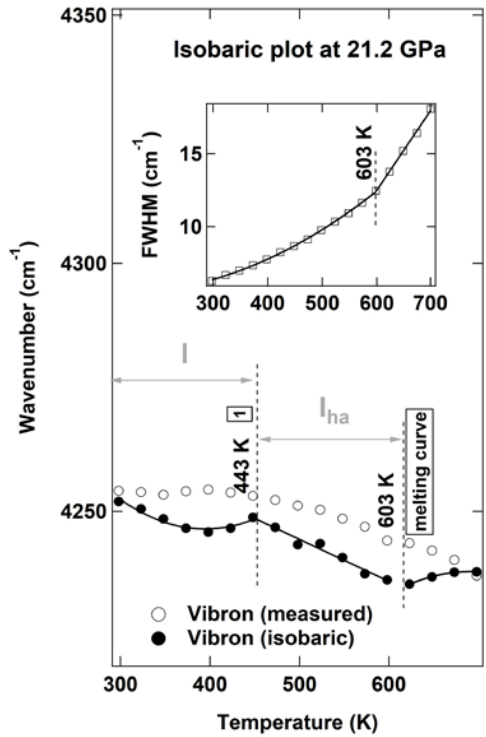

(2)

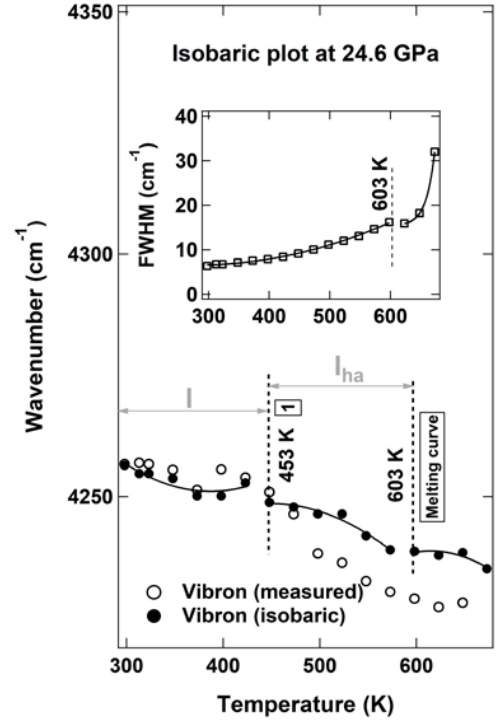

(3)

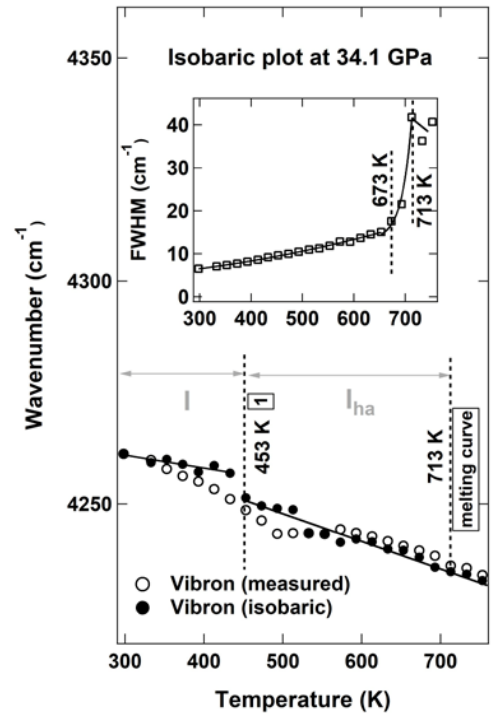

(4)
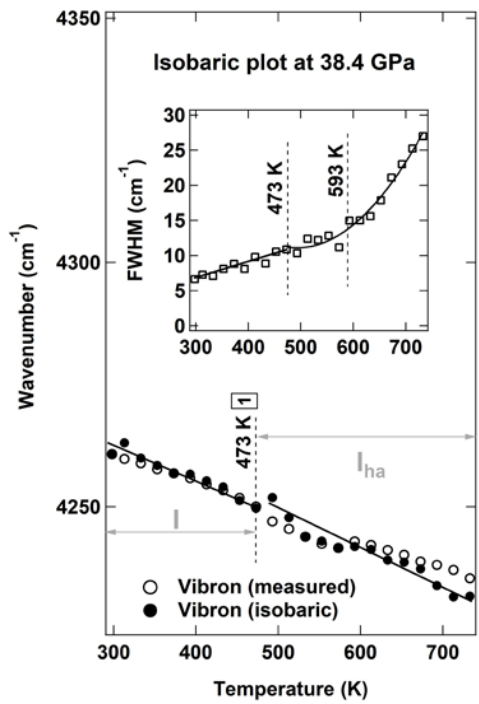

(5)
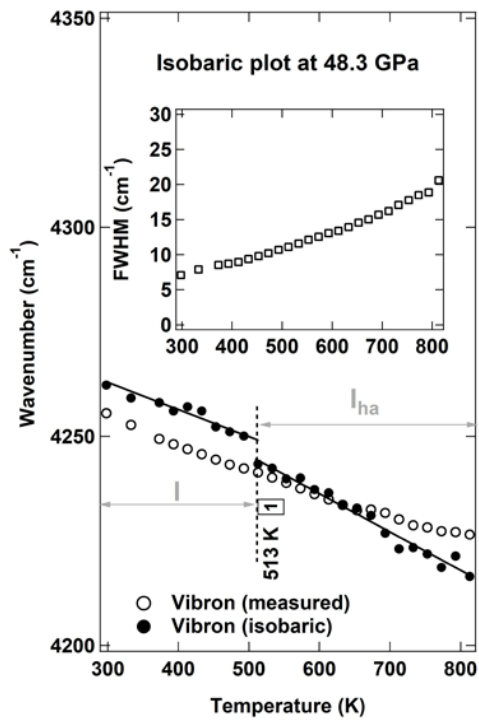

(6)
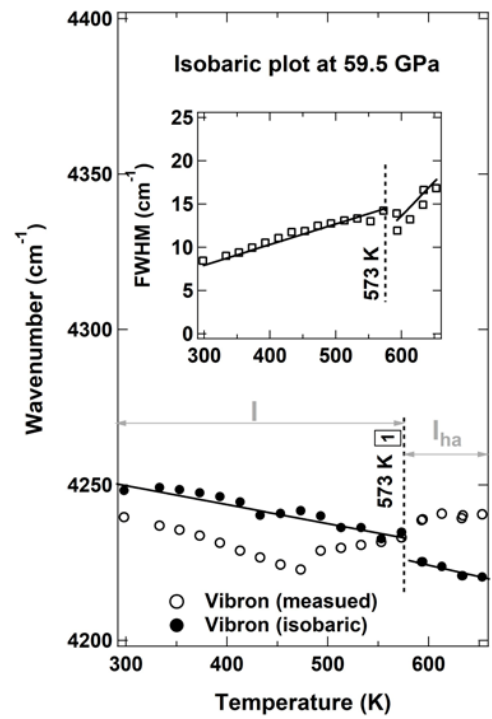

(7)
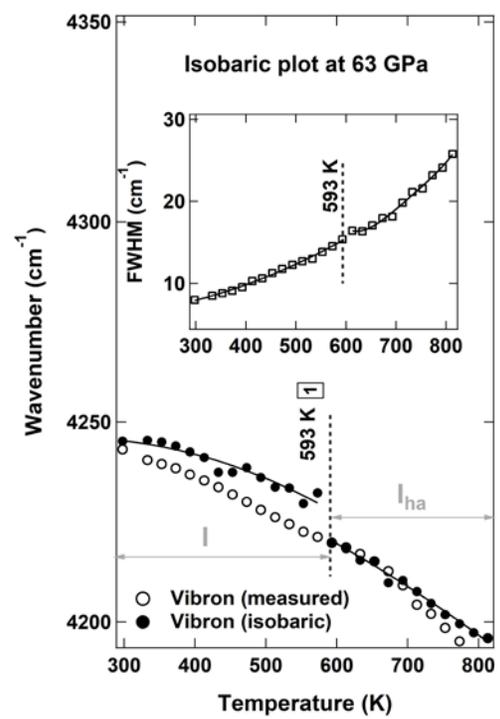

(8)
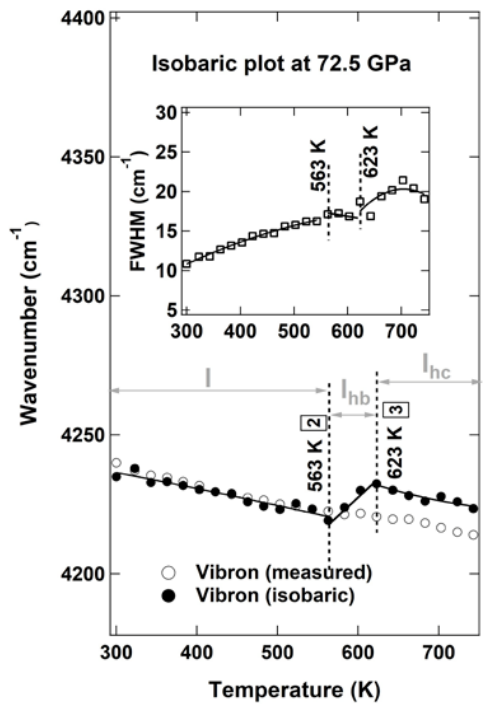

(9)
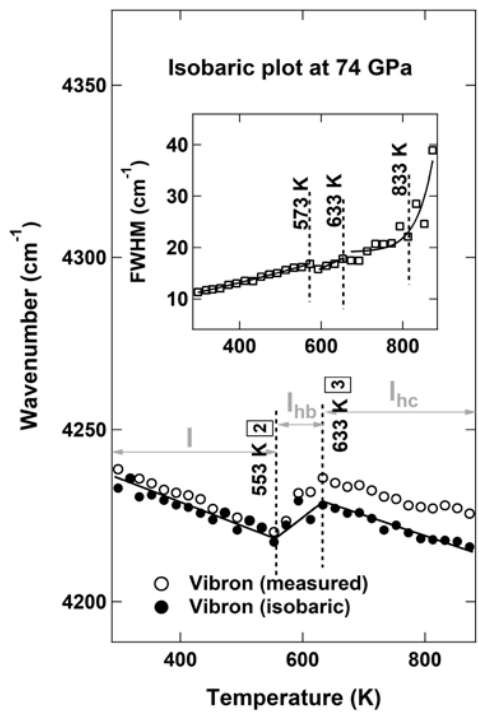

(10)
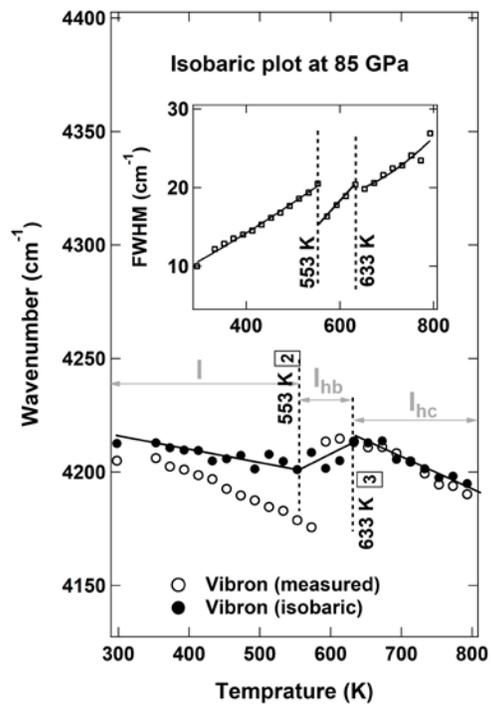

(11)
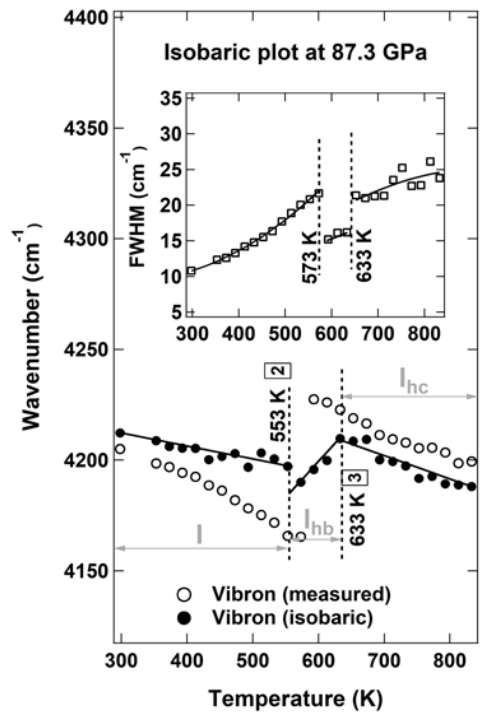

(12)

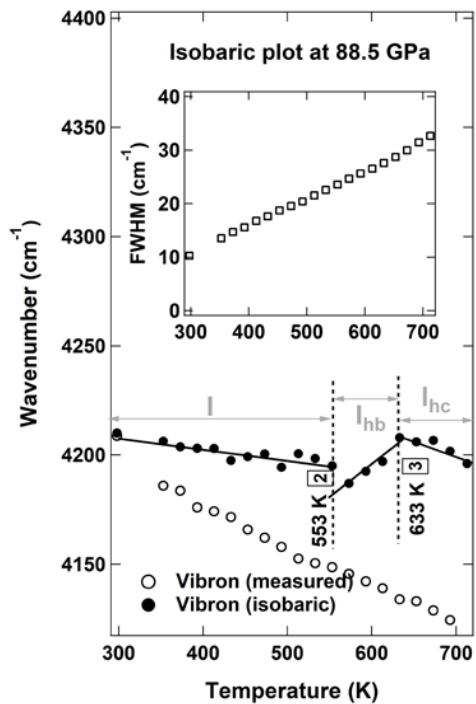

(14)

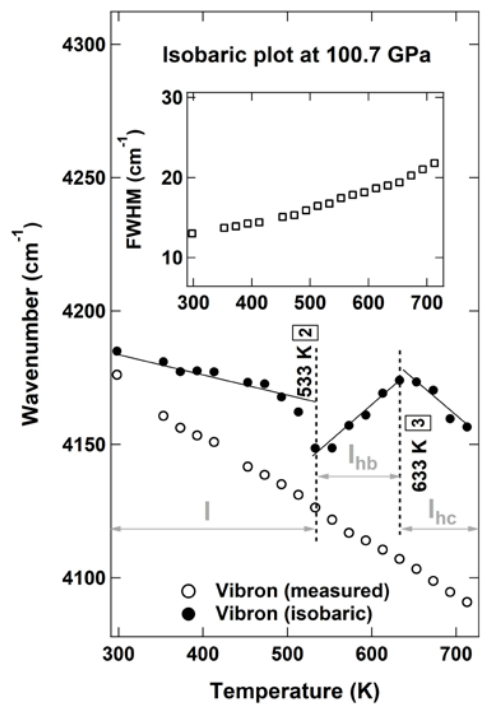

(13)

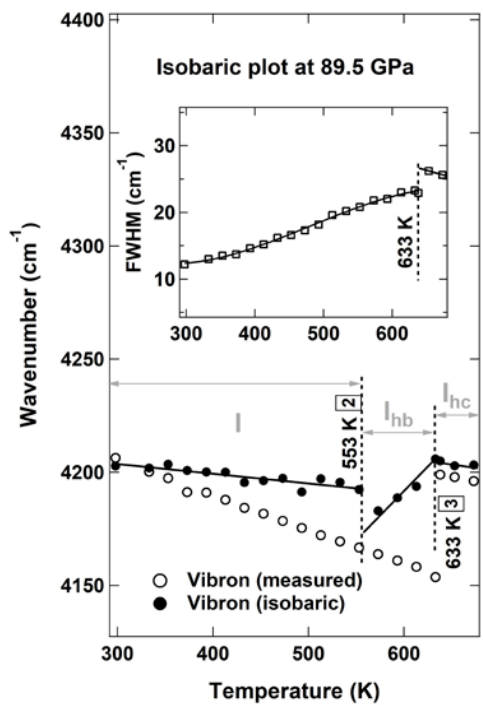

(15)

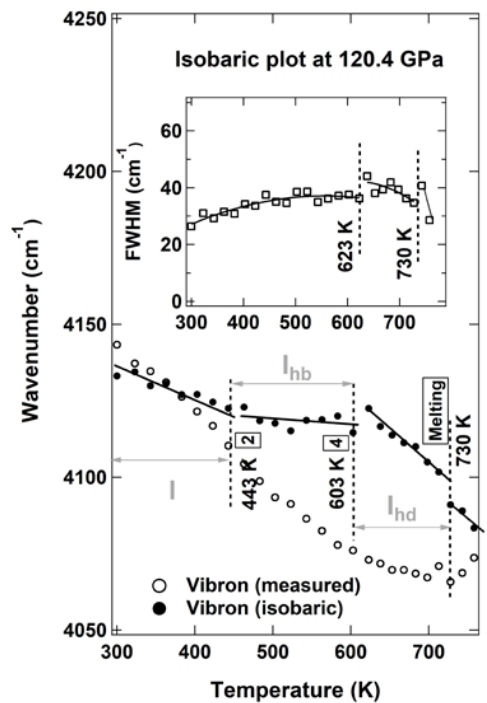

(16)
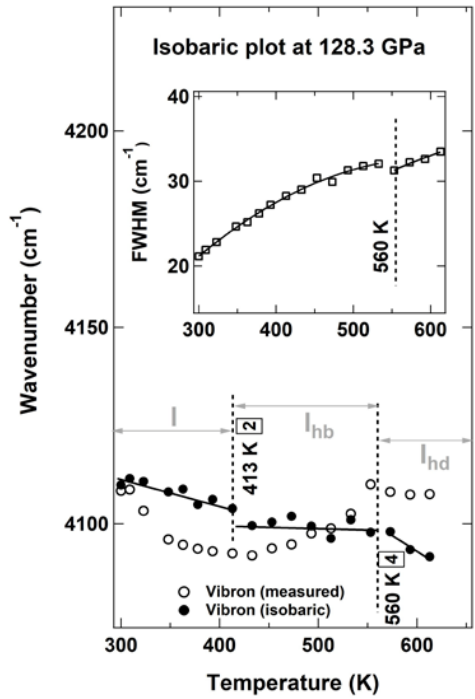

(17)
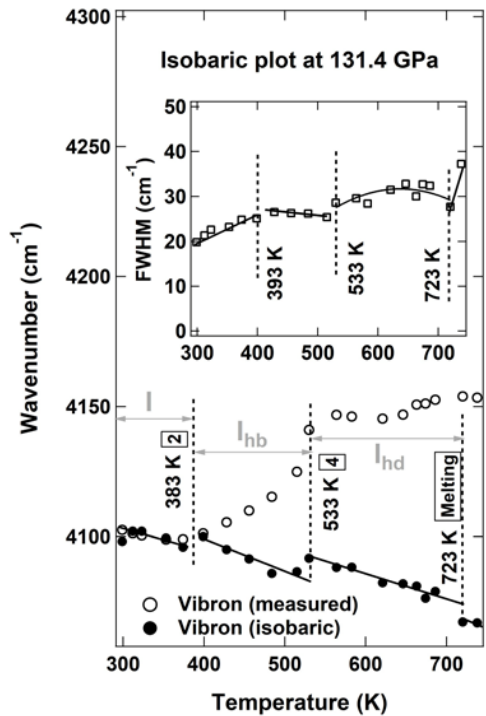

(18)
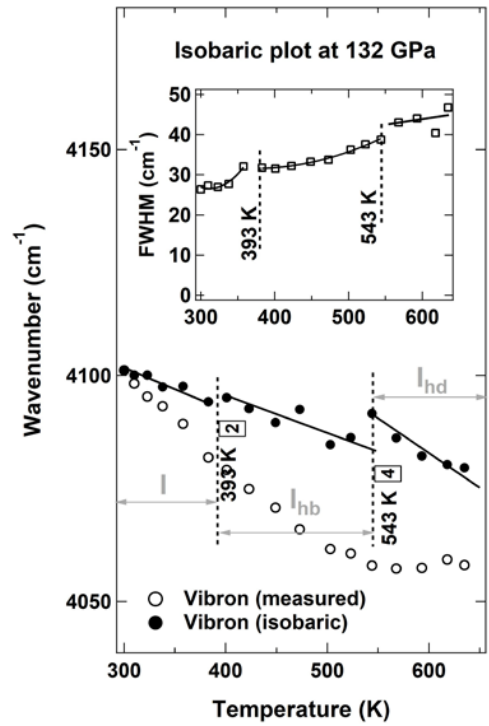

(19)
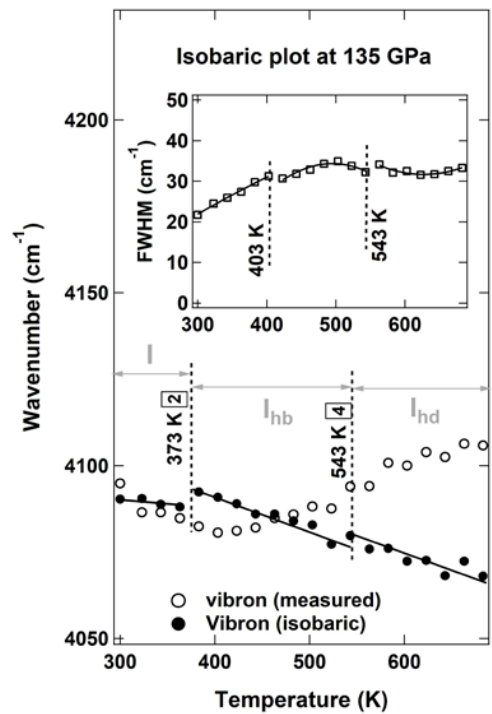

(20)
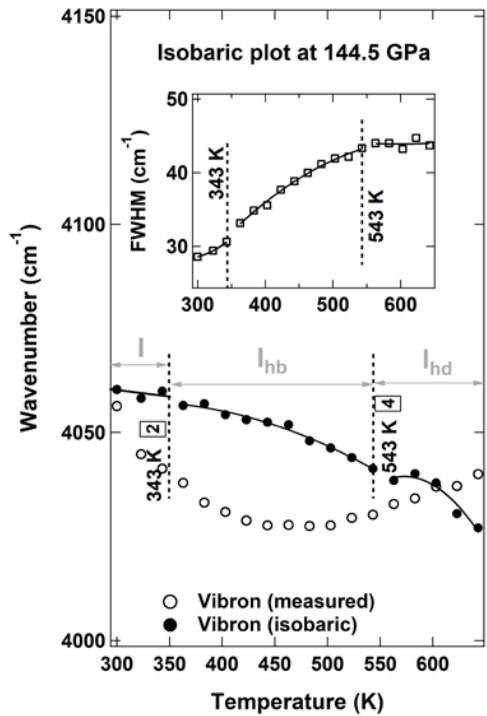

(21)
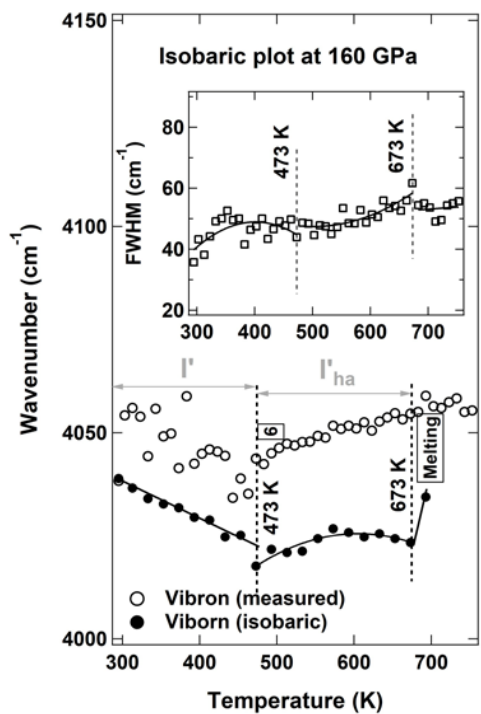

(22)
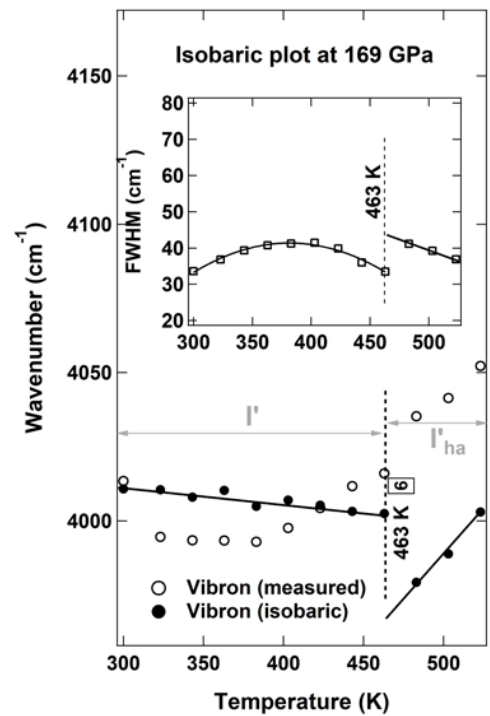

(23)
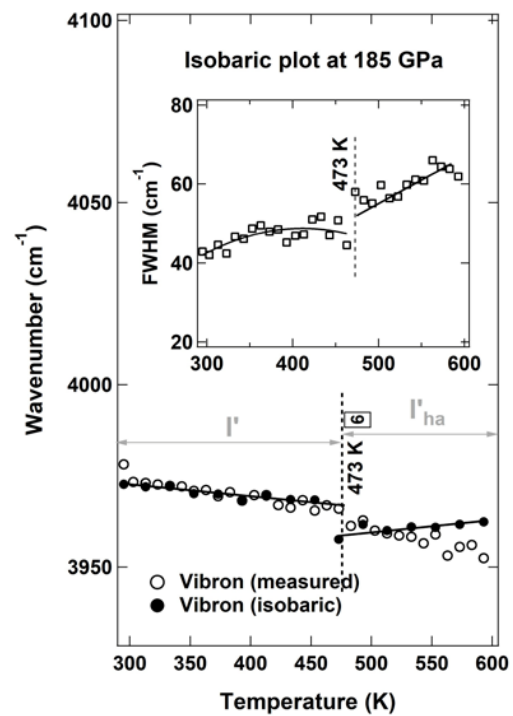

(24)
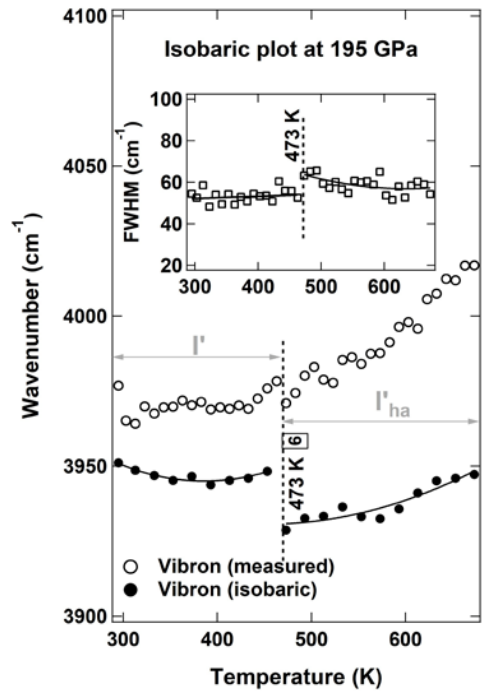

(25)
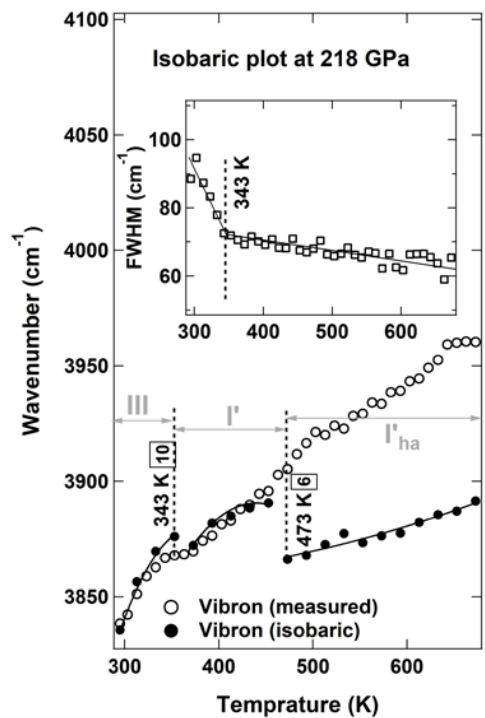

(26)
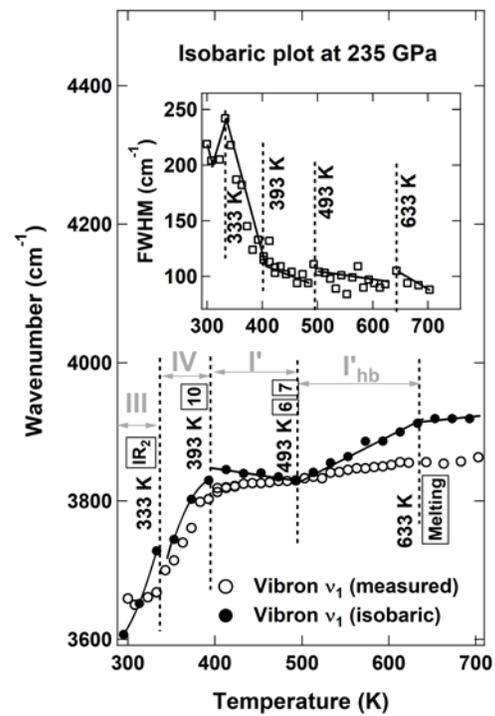

(27)
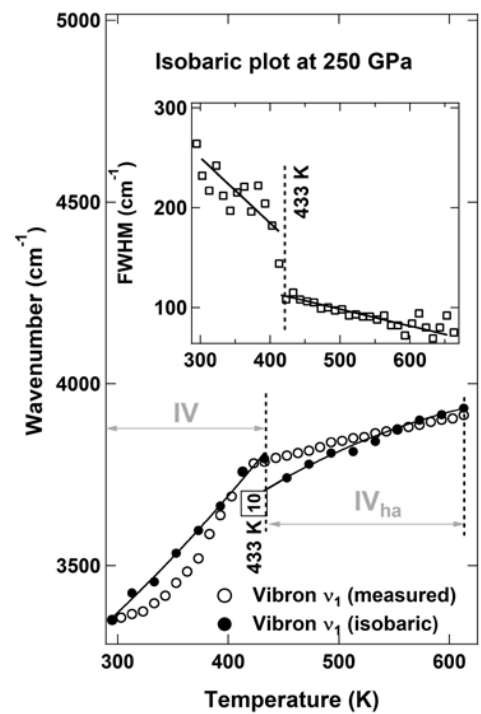

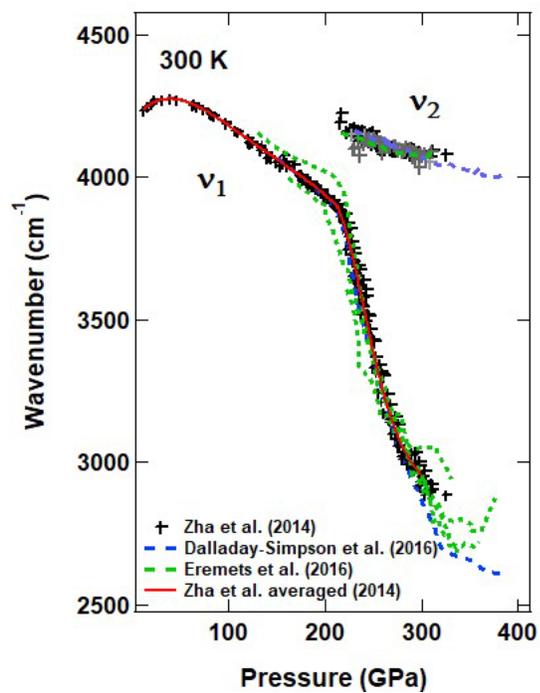

Figure S3. Pressure dependence of vibron $\nu_1$ and $\nu_2$ plotted using published data of different investigation groups.

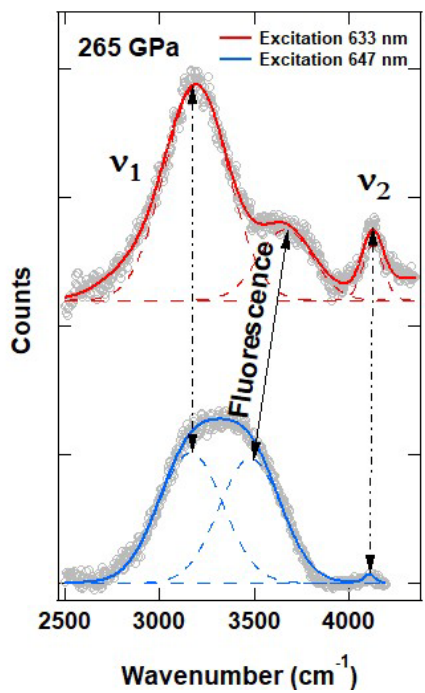

Fig. S4. Two-wavelength excitation method was used for distinguishing Raman and fluorescence peaks when pressure higher than 250 GPa.